\documentclass[10pt, conference, letterpaper]{IEEEtran}
\IEEEoverridecommandlockouts
\usepackage{cite}
\usepackage{amsmath,amssymb,amsfonts,amsthm}
\usepackage{algorithm}
\usepackage{graphicx}
\usepackage{textcomp}
\usepackage{xcolor}
\usepackage{atbegshi,picture}

\usepackage{optidef}
\usepackage{caption}
\usepackage{subcaption}
\usepackage[noend]{algpseudocode}
\usepackage{stfloats}

\def\BibTeX{{\rm B\kern-.05em{\sc i\kern-.025em b}\kern-.08em
    T\kern-.1667em\lower.7ex\hbox{E}\kern-.125emX}}
    
\usepackage[many]{tcolorbox}
\usepackage{lipsum}

\definecolor{titlebg}{RGB}{100,22,72}
\definecolor{introbg}{RGB}{0,128,128}

\newtcolorbox{usecase}[1][]{
  breakable,
  enhanced,
  arc=0pt,
  outer arc=0pt,
  colframe=titlebg,
  colback=titlebg!05,
  overlay unbroken and first={
    \node[
      draw=titlebg,
      fill=titlebg,
      rotate=0,
      anchor=north west,
      text=white,
      font=\bfseries
    ]
    at (frame.north west)  
    {#1};
  }
}

\newtcolorbox{mission}[1][]{
  breakable,
  enhanced,
  arc=0pt,
  outer arc=0pt,
  colframe=introbg,
  colback=introbg!05,
  overlay unbroken and first={
    \node[
      draw=introbg,
      fill=introbg,
      rotate=0,
      anchor=north west,
      text=white,
      font=\bfseries
    ]
    at (frame.north west)  
    {#1};
  }
}

\newcommand{\Ret}[1]{\State \textbf{return} #1}
\DeclareMathOperator*{\argmax}{arg\,max} 

\graphicspath{{images/}{figures/}}

\usepackage{xspace}
\newcommand{\FW}{\texttt{SEM-O-RAN}\xspace}

\newtheorem{theorem}{Theorem}

\usepackage[acronyms,nonumberlist,nopostdot,nomain,nogroupskip]{glossaries}

\newacronym{lidar}{LIDAR}{Light Detection and Ranging}
\newacronym{5g}{5G}{fifth-generation}
\newacronym{mmwave}{mmWave}{millimeter wave}
\newacronym{phy}{PHY}{Physical Layer}
\newacronym{mac}{MAC}{Medium Access Control}
\newacronym{uav}{UAV}{unmanned autonomous vehicle}
\newacronym{iot}{IoT}{Internet of Things}
\newacronym{ml}{ML}{machine learning}
\newacronym{drl}{DRL}{deep reinforcement learning}
\newacronym{urc}{URC}{ultra-reliable computing}
\newacronym{urllc}{URLLC}{ultra-reliable low-latency communication}

\newacronym{rfid}{RFID}{Radio Frequency Identification}
\newacronym{rfp}{RFP}{radio fingerprinting}
\newacronym{sdr}{SDR}{software-defined radio}
\newacronym{mas}{MAS}{Mobile autonomous system}
\newacronym{rl}{RL}{reinforcement learning}
\newacronym{los}{LOS}{line of sight}
\newacronym{ssm}{SSM}{spectrum sensing module}
\newacronym{dsa}{DSA}{dynamic spectrum access}
\newacronym{bpd}{BPM}{baseband processing module}
 \newacronym{ng}{NextG}{5G-and-beyond cellular networks}
\newacronym{cbrs}{CBRS}{Citizens Broadband Radio Service}
\newacronym{gaa}{GAA}{General Authorized Access}
\newacronym{pal}{PAL}{Priority Access Licensee}
\newacronym{fcc}{FCC}{Federal Communications Commission}
\newacronym{dsp}{DSP}{digital signal processing}
\newacronym{ssa}{SSA}{spectrum sensing and access}
\newacronym{fir}{FIR}{finite impulse response}
\newacronym{wsc}{WSC}{wireless signal classification}
\newacronym{ber}{BER}{bit error rate}
\newacronym{wwa}{WWI}{wideband waveform identification}
\newacronym{ppi}{PPI}{protocol and parameter inference}
\newacronym{ism}{ISM}{industrial, scientific and medical}
\newacronym{itu}{ITU}{international telecommunication union}
\newacronym{aml}{AML}{adversarial machine learning}
\newacronym{fml}{FML}{federated machine learning}
\newacronym{tml}{TML}{transfer machine learning}

\newacronym{6g}{6G}{sixth generation}
\newacronym{3gpp}{3GPP}{3rd Generation Partnership Project}
\newacronym{adc}{ADC}{Analog to Digital Converter}

\newacronym{aimd}{AIMD}{Additive Increase Multiplicative Decrease}
\newacronym{am}{AM}{Acknowledged Mode}
\newacronym{gbps}{Gbps}{Gigabit-per-second}
\newacronym{amc}{AMC}{Adaptive Modulation and Coding}
\newacronym{aqm}{AQM}{Active Queue Management}
\newacronym{awgn}{AGWN}{Additive White Gaussian Noise}
\newacronym{balia}{BALIA}{Balanced Link Adaptation}
\newacronym{bdp}{BDP}{Bandwidth-Delay Product}
\newacronym[plural=\gls{cnn}s,firstplural=convolutional neural networks (CNNs)]{cnn}{CNN}{convolutional neural network}
\newacronym[plural=\gls{dnn}s,firstplural=deep neural networks (DNNs)]{dnn}{DNN}{deep neural network}
\newacronym{bf}{BF}{Beamforming}
\newacronym{b5g}{B5G}{\gls{5g} and beyond}
\newacronym{cc}{CC}{Congestion Control}
\newacronym{cdf}{CDF}{Cumulative Distribution Function}
\newacronym{cn}{CN}{Core Network}
\newacronym{cqi}{CQI}{Channel Quality Information}
\newacronym{cp}{CP}{Control Plane}
\newacronym{csirs}{CSI-RS}{Channel State Information - Reference Signal}
\newacronym{dc}{DC}{Dual Connectivity}
\newacronym{dce}{DCE}{Direct Code Execution}
\newacronym{dci}{DCI}{Downlink Control Information}
\newacronym{dl}{DL}{deep learning}
\newacronym{dmr}{DMR}{Deadline Miss Ratio}
\newacronym{e2e}{E2E}{End-to-End}
\newacronym{ecn}{ECN}{Explicit Congestion Notification}
\newacronym{edf}{EDF}{Earliest Deadline First}
\newacronym{enb}{eNB}{evolved Node Base}
\newacronym{epc}{EPC}{Evolved Packet Core}
\newacronym{es}{ES}{Edge Server}
\newacronym{fdma}{FDMA}{Frequency Division Multiple Access}
\newacronym{fdd}{FDD}{Frequency Division Duplexing}
\newacronym[firstplural=Radio Access Technologies (RATs)]{rat}{RAT}{Radio Access Technology}
\newacronym{fs}{FS}{Fast Switching}
\newacronym{ftp}{FTP}{File Transfer Protocol}
\newacronym{gnb}{gNB}{Next Generation Node Base}
\newacronym{harq}{HARQ}{Hybrid Automatic Repeat reQuest}
\newacronym{hetnet}{HetNet}{Heterogeneous Network}
\newacronym{hh}{HH}{Hard Handover}
\newacronym{hol}{HOL}{Head-of-Line}
\newacronym{ia}{IA}{Initial Access}
\newacronym{imt}{IMT}{International Mobile Telecommunication}
\newacronym{lte}{LTE}{Long Term Evolution}
\newacronym{m2m}{M2M}{Machine to Machine}
\newacronym{mc}{MC}{Multi-Connectivity}
\newacronym{mcs}{MCS}{Modulation and Coding Scheme}
\newacronym{mi}{MI}{Mutual Information}
\newacronym{mptcp}{MPTCP}{Multipath TCP}
\newacronym{mr}{MR}{Maximum Rate}
\newacronym{mss}{MSS}{Maximum Segment Size}
\newacronym{mtd}{MTD}{Machine-Type Device}
\newacronym{mtu}{MTU}{Maximum Transmission Unit}
\newacronym{nfv}{NFV}{Network Function Virtualization}
\newacronym{nlos}{NLOS}{Non-Line-of-Sight}
\newacronym{nr}{NR}{New Radio}
\newacronym{ofdm}{OFDM}{Orthogonal Frequency Division Multiplexing}
\newacronym{pdcch}{PDCCH}{Physical Downlink Control Channel}
\newacronym{pdcp}{PDCP}{Packet Data Convergence Protocol}
\newacronym{pdsch}{PDSCH}{Physical Downlink Shared Channel}
\newacronym{pdu}{PDU}{Packet Data Unit}
\newacronym{pf}{PF}{Proportional Fair}
\newacronym{pgw}{PGW}{Packet Gateway}
\newacronym{pbch}{PBCH}{Physical Broadcast Channel}
\newacronym[plural=\gls{mme}s,firstplural=Mobility Management Entities (MMEs)]{mme}{MME}{Mobility Management Entity}
\newacronym{prb}{PRB}{Physical Resource Block}
\newacronym{pss}{PSS}{Primary Synchronization Signal}
\newacronym{pu}{PU}{primary users}
\newacronym{su}{SU}{secondary user}
\newacronym{pucch}{PUCCH}{Physical Uplink Control Channel}
\newacronym{pusch}{PUSCH}{Physical Uplink Shared Channel}
\newacronym{rach}{RACH}{Random Access Channel}
\newacronym{red}{RED}{Random Early Detection}
\newacronym{rf}{RF}{Radio Frequency}
\newacronym{rlc}{RLC}{Radio Link Control}
\newacronym{rlf}{RLF}{Radio Link Failure}
\newacronym{rrc}{RRC}{Radio Resource Control}
\newacronym{rrm}{RRM}{Radio Resource Management}
\newacronym{rr}{RR}{Round Robin}
\newacronym{rs}{RS}{Remote Server}
\newacronym{rsrp}{RSRP}{Reference Signal Received Power}
\newacronym{rss}{RSS}{Received Signal Strength}
\newacronym{rtt}{RTT}{Round Trip Time}
\newacronym{rw}{RW}{Receive Window}
\newacronym{rx}{RX}{Receiver}
\newacronym{sa}{SA}{standalone}
\newacronym{sack}{SACK}{Selective Acknowledgment}
\newacronym{sap}{SAP}{Service Access Point}
\newacronym{sch}{SCH}{Secondary Cell Handover}
\newacronym{scoot}{SCOOT}{Split Cycle Offset Optimization Technique}
\newacronym{fpga}{FPGA}{field-programmable gate array}
\newacronym{sdma}{SDMA}{Spatial Division Multiple Access}
\newacronym{sinr}{SINR}{Signal to Interference plus Noise Ratio}
\newacronym{sm}{SM}{Saturation Mode}
\newacronym{snr}{SNR}{Signal-to-Noise-Ratio}
\newacronym{son}{SON}{Self-Organizing Network}
\newacronym{ss}{SS}{Synchronization Signal}
\newacronym{srs}{SRS}{Sounding Reference Signal}
\newacronym{sss}{SSS}{Secondary Synchronization Signal}
\newacronym{tb}{TB}{Transport Block}
\newacronym{tcp}{TCP}{Transmission Control Protocol}
\newacronym{tdd}{TDD}{Time Division Duplexing}
\newacronym{tdma}{TDMA}{Time Division Multiple Access}
\newacronym{tfl}{TfL}{Transport for London}
\newacronym{tm}{TM}{Transparent Mode}
\newacronym{trp}{TRP}{Transmitter Receiver Pair}
\newacronym{tti}{TTI}{Transmission Time Interval}
\newacronym{ttt}{TTT}{Time-to-Trigger}
\newacronym{tx}{TX}{Transmitter}
\newacronym{ue}{UE}{User Equipment}
\newacronym{ul}{UL}{Uplink}
\newacronym{uml}{UML}{Unified Modeling Language}
\newacronym{um}{UM}{Unacknowledged Mode}
\newacronym{utc}{UTC}{Urban Traffic Control}
\newacronym{vm}{VM}{Virtual Machine}
\newacronym{rsrq}{RSRQ}{Reference Signal Received Quality}
\newacronym{rssi}{RSSI}{Received Signal Strength Indicator}
\newacronym{crs}{CRS}{Cell Reference Signal}
\newacronym{nsa}{NSA}{Non Stand Alone}
\newacronym{mrdc}{MR-DC}{Multi \gls{rat} \gls{dc}}
\newacronym{endc}{EN-DC}{E-UTRAN-\gls{nr} \gls{dc}}
\newacronym{5gc}{5GC}{5G Core}
\newacronym{si}{SI}{Study Item}
\newacronym{iab}{IAB}{Integrated Access and Backhaul}
\newacronym{wf}{WF}{Wired-first}
\newacronym{hqf}{HQF}{Highest-quality-first}
\newacronym{pa}{PA}{Position-aware}
\newacronym{mlr}{MLR}{Maximum-local-rate}
\newacronym{wbf}{WBF}{Wired Bias Function}
\newacronym{mib}{MIB}{Master Information Block}
\newacronym{sib}{SIB}{Secondary Information Block}
\newacronym{kpi}{KPI}{Key Performance Indicator}
\newacronym{ppp}{PPP}{Poisson Point Process}
\newacronym{gtp}{GTP}{GPRS Tunneling Protocol}
\newacronym{amf}{AMF}{Access and Mobility Management Function}
\newacronym{dash}{DASH}{Dynamic Adaptive Streaming over HTTP}
\newacronym{http}{HTTP}{HyperText Transfer Protocol}
\newacronym{qos}{QoS}{Quality of Service}
\newacronym{udp}{UDP}{User Datagram Protocol}
\newacronym{mt}{MT}{Mobile Termination}
\newacronym{sdap}{SDAP}{Service Data Adaptation Protocol}
\newacronym{tdm}{TDM}{Time Division Multiplexing}
\newacronym{fdm}{FDM}{Frequency Division Multiplexing}
\newacronym{sdm}{SDM}{Space Division Multiplexing}
\newacronym{dag}{DAG}{Directed Acyclic Graph}
\newacronym{st}{ST}{Spanning Tree}
\newacronym{ummimo}{UM-MIMO}{Ultra-massive Multiple Input, Multiple Output}
\newacronym{wlan}{WLAN}{Wireless LAN}
\newacronym{wlans}{WLANs}{Wireless Local Area Networks}
\newacronym{rlnc}{RLNC}{Random Linear Network Coding}
\newacronym{drx}{DRX}{Discontinuous Reception}
\newacronym{cpu}{CPU}{Central Processing Unit}
\newacronym{soc}{SoC}{system-on-chip}

\newacronym{ap}{AP}{Access Point}
\newacronym{dfx}{DFX}{dynamic function exchange}
\newacronym{cv}{CV}{computer vision}

\newacronym{mimo}{MIMO}{multiple-input multiple-output}
\newacronym{mu}{MU}{Multi-user}
\newacronym{mumimo}{MU-MIMO}{\gls{mu}-\gls{mimo}}
\newacronym{dcm}{DCM}{distributed cooperative \gls{mimo}}
\newacronym{comp}{CoMP}{Coordinated Multi-Point}

\newacronym{v2x}{V2X}{vehicle-to-everything}

\newacronym{rb}{RB}{resource block}
\newacronym{vno}{VNO}{Virtual Network Operator}

\newacronym{ran}{RAN}{Radio Access Network}
\newacronym{oran}{O-RAN}{Open \gls{ran}}
\newacronym{ric}{RIC}{RAN Intelligent Controller}

\newcommand{\rt}{Near-real-time \gls{ric}\xspace}
\newcommand{\nrt}{Non-real-time \gls{ric}\xspace}

\newacronym{cu}{CU}{Centralized Unit}
\newacronym{du}{DU}{Distributed Unit}
\newacronym{ru}{RU}{Radio Unit}

\newacronym{ets}{ETS}{expeditionary tactical system}
\newacronym{cccc}{4C}{Congested, Contested, Concealed, and Contaminated}
\newacronym{swap}{SWaP}{Size, Weight, and Power}

\newacronym{rsd}{RSD}{Radiofrequency Spectrum Dominance}
\newacronym{iobt}{IoBT}{Internet of Battlefield Things}

\newacronym{sdla}{SDLA}{Semantic Deep Learning Analyzer}
\newacronym{sesm}{SESM}{Semantic Edge Slicing Module}
\newacronym{sfesp}{SF-ESP}{Semantic Flexible Edge Slicing Problem}
\newacronym{td}{TD}{Task Description}
\newacronym{tr}{TR}{Task Requirements}
\newacronym{osr}{OSR}{O-RAN Slice Request}
\newacronym{rbg}{RBG}{Resource Block Group}
\newacronym{mAP}{mAP}{mean Average Precision}

\newacronym{lpwan}{LPWAN}{low-power wide-area network}
\newacronym{sf}{SF}{Spreading Factor}
\newacronym{ai}{AI}{Artificial Intelligence}
\newacronym{nbiot}{NB-IoT}{Narrowband IoT}
\newacronym{fl}{FL}{Federated Learning}
\newacronym{arq}{ARQ}{Automatic Repeat Request}
\newacronym{fec}{FEC}{Forward Error Correction}
\newacronym{minlp}{MINLP}{Mixed-Integer Nonlinear Programming}
\newacronym{srn}{SRN}{Standard Radio Node}
\newacronym{mchem}{MCHEM}{Massive Channel Emulator}
\newacronym{mec}{MEC}{mobile edge computing}
    
\AtBeginShipoutNext{\AtBeginShipoutUpperLeft{%
  \put(16.1cm,-0.5cm){\makebox[0pt][r]{This paper has been accepted in IEEE INFOCOM 2023 Main Conference}}%
}}
\begin{document}

\title{\FW: Semantic and Flexible O-RAN Slicing for NextG Edge-Assisted Mobile Systems \vspace{-0.2cm}
}

\author{\IEEEauthorblockN{Corrado Puligheddu$^{\dagger}$, Jonathan Ashdown$^{\circ}$, Carla Fabiana Chiasserini$^{\dagger}$, and Francesco Restuccia$^{\ddagger}$\vspace{-0.5cm}}\\
\IEEEauthorblockA{
$\dagger$ Politecnico di Torino, Italy\\
$\circ$ Air Force Research Laboratory, United States\\
$\ddagger$ Institute for the Wireless Internet of Things, Northeastern University, United States
}
\vspace{-1.2cm}
\thanks{
\copyright 2022 IEEE. Personal use of this material is permitted. Permission from IEEE must be
obtained for all other uses, in any current or future media, including
reprinting/republishing this material for advertising or promotional purposes, creating new
collective works, for resale or redistribution to servers or lists, or reuse of any copyrighted
component of this work in other works.

Approved for Public Release; Distribution Unlimited: AFRL-2022-1622}
}

\maketitle

\thispagestyle{plain}
\pagestyle{plain}
\pagenumbering{gobble}

\begin{abstract}
5G and beyond cellular networks (NextG)  will support the continuous execution of resource-expensive edge-assisted \gls{dl} tasks. To this end, \gls{ran} resources will need to be carefully ``sliced'' to satisfy heterogeneous application requirements while minimizing \gls{ran} usage. Existing slicing frameworks treat each DL task as equal and inflexibly define the resources to assign to each task, which leads to sub-optimal performance. In this paper, we propose \FW, the first \textit{semantic} and \emph{flexible} slicing framework for NextG Open RANs. Our key intuition is that different \gls{dl} classifiers can tolerate different levels of image compression, due to the semantic nature of the target classes. Therefore, compression can be semantically applied so that the networking load can be minimized. Moreover, flexibility allows \FW to consider multiple edge allocations leading to the same task-related performance, which significantly improves system-wide performance as more tasks can be allocated. First, we mathematically formulate the Semantic Flexible Edge Slicing Problem (SF-ESP), demonstrate that it is NP-hard, and provide an approximation algorithm to solve it efficiently. Then, we  evaluate the performance of \FW through extensive numerical analysis with state-of-the-art multi-object detection (YOLOX) and image segmentation (BiSeNet V2), as well as real-world experiments on the Colosseum testbed. Our results show that \FW improves the number of allocated tasks by up to 169\% with respect to the state of the art.
\end{abstract}

\vspace{-0.2cm}
\section{Introduction}
The number of mobile devices using \gls{ng} is expected to reach 64 billion by 2025 \cite{EricssonMobilityReport2021}. Among others, \gls{v2x} communications \cite{chen2017vehicle,zugno2020toward} are  enabling autonomous driving \cite{bagheri20215g} and drone-based delivery \cite{frachtenberg2019practical}.  Thanks to \gls{v2x}, the self-driving car market will reach global revenue of \$49.79B by 2024 \cite{SelfDriving}. 

To perform their mission-critical operations, \gls{v2x} and other mobile devices will continuously execute complex \gls{cv}-based \gls{dl} tasks, which require as input high-resolution images (e.g., frames of a video) or three-dimensional LIDAR (Light Detection and Ranging) data \cite{hao2018machine}. Examples include multi-object classification of  blockages, intersections, driveways, fire hydrants, and people \cite{ravindran2020multi}. However,  continuously sending multimedia data to the edge may eventually saturate the \gls{ran}. For example, in the Cityscape dataset \cite{cordts2016cityscapes}, images have a 100 KB size on average. By assuming that real-time self-navigation requires DL inference on frames collected from 4 cameras each 10 ms, the traffic load would be 32 Gb/s if 100 vehicles are connected to the \gls{ran}. 


To this end, \gls{ran} slicing \cite{li20215growth,d2020sl,mandelli2019satisfying,mancuso2019slicing,d2019slice,garcia2018posens} allows \glspl{vno} to virtualize and allocate the computational and networking resources of the \gls{ran} according to their needs. Interestingly, \gls{ran} slicing is fully supported by the \gls{oran} framework, which disaggregates the \gls{ng} \gls{ran} hardware from its software components to allow fine-grained real-time flexible control of the \gls{ran} components \cite{polese2022understanding,doro2022orchestran,bonati2021intelligence}, as summarized in Section \ref{sec:oran}.


\textbf{Existing Issues.}~The current state of the art -- discussed in detail in Section \ref{sec:rw} -- either does not support \gls{oran} or defines edge-based tasks in a \textit{monolithic} fashion, which leads to sub-optimal performance, as shown in Section \ref{sec:numres}. To this end, we propose \FW, the first \gls{oran} slicing framework for \gls{ng} edge-assisted mobile applications. 

\begin{figure}
  \centering
  \begin{subfigure}{.49\linewidth}
  \includegraphics[width=\linewidth,height=3.5cm,keepaspectratio]{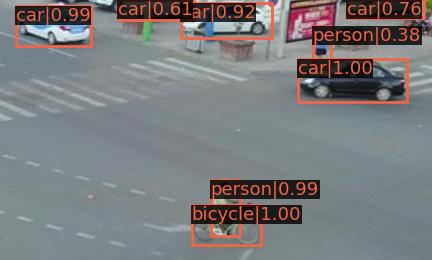}%
  \caption{Compression 0.47x, 29.5 KB}
  \label{sub:comp_light}
  \end{subfigure}
  \begin{subfigure}{.49\linewidth}
  \includegraphics[width=\linewidth,height=3.5cm,keepaspectratio]{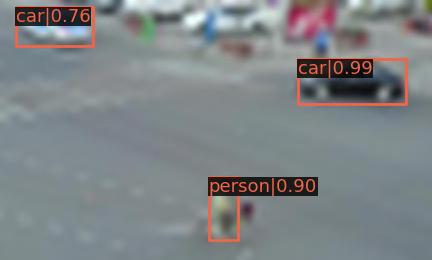}%
  \caption{Compression 0.04x, 2.3~KB}
  \label{sub:comp_heavy}
  \end{subfigure}
  \caption{Stronger compression rates make some objects undetectable and/or harder to detect by CV-based DL models.\vspace{-0.2cm}}
  \label{fig:compression}
  \vspace{-0.4cm}
\end{figure}

\textbf{Two core innovations separate \FW from the state of the art}. First, existing work \textit{pre-defines} the number and type of edge resources needed to perform a given task. Conversely, we define a task in terms of \textit{required end-to-end latency and accuracy-per-class performance}, thus allowing \emph{flexibility} in the way edge resources are allocated. Flexibility allows for the consideration of multiple edge allocations leading to the same task-related performance, ultimately improving system-wide performance. In Section \ref{sec:experiments}, we show that flexibility improves the number of allocated tasks by up to \textbf{31}\% with respect to the state of the art \cite{d2020sl}. Second, \FW considers the \emph{semantics} of the DL task to further reduce the network overhead by compressing the images. For example, Fig.~\ref{fig:compression} shows that classifying cars is semantically less difficult than bicycles, thus images can be compressed more aggressively if classifying cars is the priority. In Section \ref{sec:experiments}, we show that combining flexibility and semantics improves the performance by up to \textbf{169\%} with respect to \cite{d2020sl}.  \smallskip


\textbf{Technical Challenges.}~Introducing flexibility and application semantics into the O-RAN slicing mathematical formulation is challenging, since \textbf{(i)} the relationship between the allocated slice, image compression, classification accuracy for the target classes, and network latency  cannot be easily expressed in closed form, since state-of-the-art DL models are highly non-linear; \textbf{(ii)} the flexibility in edge resource allocation makes the optimization significantly more complex, as shown in Section \ref{sec:sesp_formulation}. \textit{To the best of our knowledge, no other work has holistically tackled these two aspects at the same time}. \vspace{-0.2cm}

\subsection*{\textbf{Summary of Novel Contributions}}

$\bullet$ We present \FW, the first \textit{semantic} and \textit{flexible}  slicing framework to support  edge-assisted DL task offloading in NextG networks. \FW is fully compliant with the O-RAN specifications (Section \ref{sec:framework}), which allow for the near-real-time control of slices configuration. To perform the actual slicing, we mathematically formulate the Semantic Flexible Edge Slicing Problem (SF-ESP), which (i) optimizes the number of DL tasks executed at the RAN edge while (ii) guaranteeing strict guarantees on the DL task latency/accuracy, and (iii) avoiding resource over-provisioning (Section \ref{sec:semproblem}). \textbf{The SF-ESP is fundamentally different from existing formulations}, since (a) it incorporates highly non-linear relationships between slicing, compression, end-to-end latency, and classification accuracy; (b) employs flexibility in resource assignments to balance the consumption of the different types of resources and avoid the depletion of the most requested ones. We demonstrate that the SF-ESP is NP-hard, and propose a greedy algorithm to solve it efficiently (Section \ref{sec:approximate}); \vspace{0.1cm}


$\bullet$ We evaluate \FW through extensive numerical analysis (Section \ref{sec:numres}) and through a prototype implemented on the Colosseum network emulator \cite{bonati2021colosseum} (Section \ref{sec:experiments}). We consider two state-of-the-art \gls{cv} problems, i.e., multi-object detection with the YOLOX model \cite{ge2021yolox} and the COCO dataset \cite{lin2014microsoft}, as well as the image segmentation problem on the Cityscapes urban mobility dataset \cite{cordts2016cityscapes} with the BiSeNet v2 real-time classifier \cite{Yu2021}. We compare \FW with 5 baselines, including the state-of-the-art Sl-EDGE framework \cite{d2020sl}. Our results show that \FW improves the number of allocated tasks by up to 169\% and by 18\% on average with respect to Sl-EDGE. \textbf{To allow replicability and benchmarking, we have released our algorithm as open-source\footnote{https://github.com/corrado113/Semoran}.}
\vspace{-0.2cm}

\section{Two Key Concepts in \FW}\label{sec:sem_idea}

\begin{figure}[!h]
    \centering
    \begin{subfigure}[t]{0.4575\columnwidth}
    \centering
    \includegraphics[width=\columnwidth]{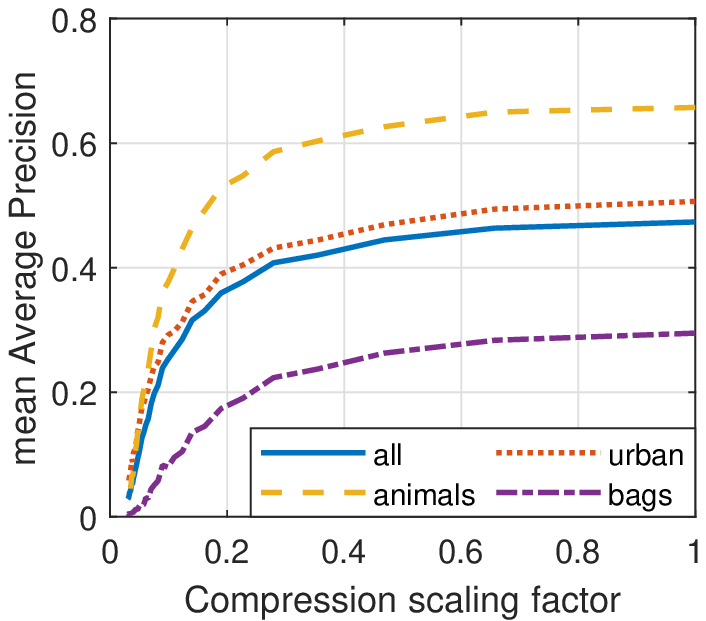}
    \end{subfigure}
    \begin{subfigure}[t]{0.5225\columnwidth}
    \centering
    \includegraphics[width=\columnwidth]{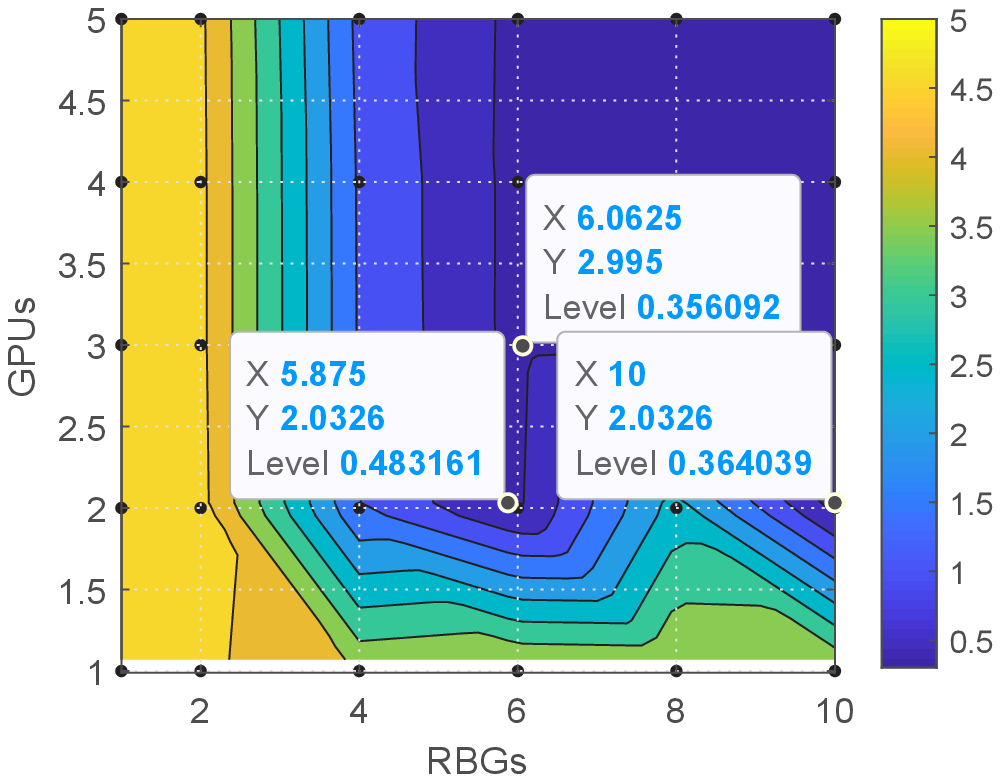}
    \end{subfigure}
    \caption{(Left) Mean Average Precision (mAP) as a function of the compression scaling factor for the application classes defined in Tab.~\ref{tab:applications}; (Right) Experimental latency as a function of allocated radio Resource Block Groups (RBGs) and GPUs.\vspace{-0.3cm}}
    \label{fig:semantic_allocation}
\end{figure}

The first main concept is the semantic-based slicing. In Fig. \ref{fig:compression}, we notice that different target classes have different tolerances to image compression. Intuitively, some classes are semantically "harder" than others, especially in some circumstances. For example, a person or a car can be more easily identified in a noisy image as opposed to a bicycle or a backpack. In the left side of Fig.~\ref{fig:semantic_allocation} we quantitatively evaluate this behaviour, showing the \gls{mAP} values corresponding to the different mobile sensing applications defined in Tab. \ref{tab:applications}. The \gls{mAP} is a metric used to evaluate object detection models, defined as the mean over all object classes of the area under the Precision-Recall Curve.  \textit{The takeaway point here is that there is a margin for significant compression on the images sent to the edge for inference, while still obtaining acceptable inference accuracy on average}.  

The second concept is the flexibility in task resource allocation. Indeed, a task requires many different kinds of resources, from networking to computation and storage. Therefore, the slicing algorithm can allocate  different amounts of resources in each category and still meet performance requirements. To illustrate this point, the right side of Fig.~\ref{fig:semantic_allocation} shows experimental end-to-end task latency results of inference on the state-of-the-art YOLOX \gls{dnn} model for object detection \cite{ge2021yolox} computed using the Colosseum network emulator \cite{bonati2021colosseum}, as a function of the allocated \glspl{rbg} and GPUs. In this plot, 10 images per second were generated from a single \gls{ue}, without employing image compression. 

\begin{figure*}[!h]
  \centering
  \includegraphics[width=\linewidth]{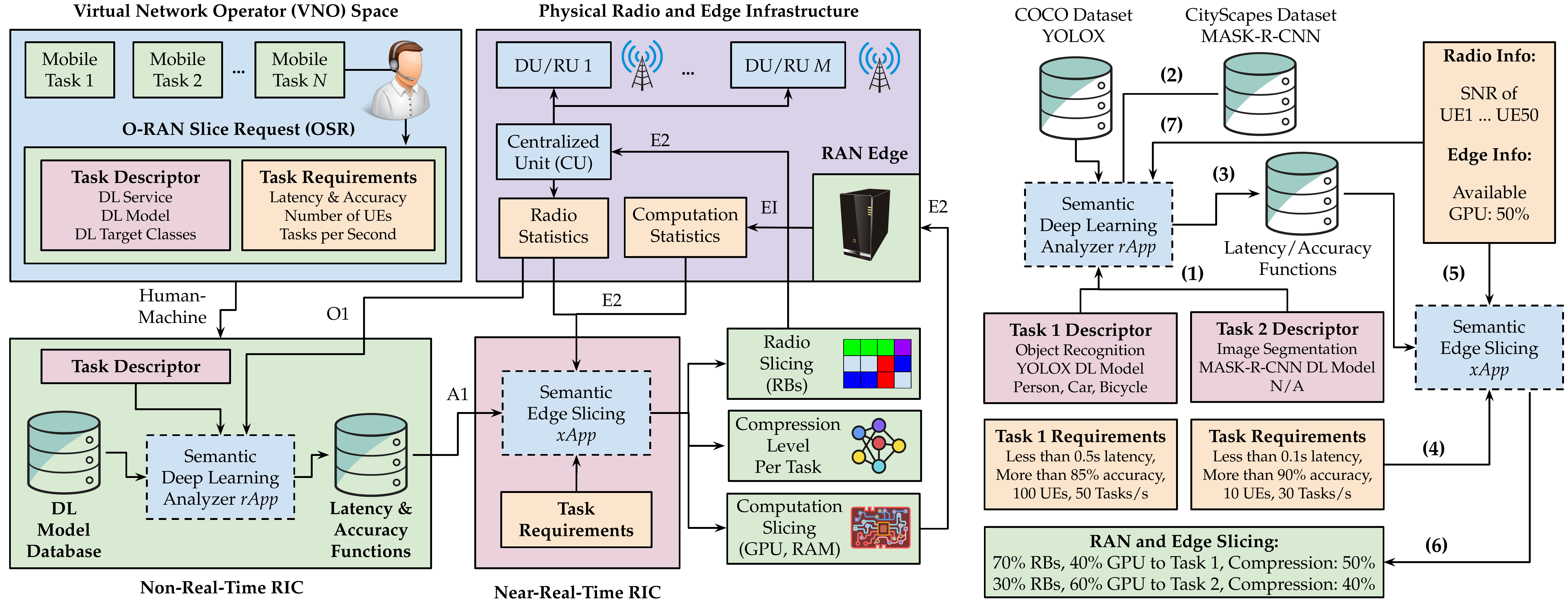}
  \caption{Functional Blocks and O-RAN Interfaces used by \FW (Left); A Walk-through of \FW (Right).\vspace{-0.3cm}}
  \label{fig:framework_oran}
\end{figure*}

\textit{The key takeaway is that more than one combination of RBG/GPU allocations can lead to the same latency performance while allowing more allocated tasks}. For example, let us assume that 25 RGBs and 4 GPUs are the maximum radio and computational resources available in the RAN, and that two tasks (T1 and T2) requiring 0.4 s of latency need to be allocated. According to Fig.~\ref{fig:semantic_allocation}, two different RGB/GPU allocations meet the 0.4s latency requirement, namely (6, 3) and (10, 2). Let us assume T1 is allocated (6, 3), which is the most resource-efficient allocation. In this case, however, T2 could not be allocated as there would only be 1 GPU  left. Instead, if (10, 2) is allocated to T1, T2 can be allocated since 2 GPUs and 10 RBGs are still available. 

 

\section{The \FW Framework}\label{sec:framework}


\subsection{Background Notions on O-RAN}\label{sec:oran}

The core philosophy behind \gls{oran} is the clear separation the RAN software and hardware \cite{bonati2020open}, by disaggregating the RAN into a \gls{ru}, \gls{cu} and \gls{du}. The RU implements extremely low-latency operations related to the lower \gls{phy}. The DU, in turn, implements the upper portion of the \gls{phy}, as well as the \gls{mac} and \gls{rlc}. These are controlled in a softwarized manner by a \gls{ric}, which is further divided into a \nrt, handling high-level \gls{ran} orchestration and management, and a \rt, implementing fine-grained control policies such as \gls{ran} slicing, scheduling, and load balancing.  Third-party applications called xApps and rApps can be hosted in the \nrt and \rt, respectively. The former may implement data-driven control loops or may be used for \gls{ran}-specific data collection and analysis. On the other hand, rApps may implement high-level policy guidance as well as application-level interfaces. Please refer to \cite{polese2022understanding} for more information regarding O-RAN.


\subsection{\FW: Functional Blocks and Interfaces}\label{sec:sem_blocks}

Fig.~\ref{fig:framework_oran} shows the functional blocks of \FW, as well as how the blocks are mapped into the O-RAN modules and interfaces. The core modules of \FW are the \gls{sdla} and the \gls{sesm}, which respectively reside in the \nrt and \rt portions of the O-RAN as an rApp and an xApp. The \FW and the \gls{vno} communicate through a human-machine interface \cite{polese2022understanding}. Each \gls{vno} requires slices for a given set of mobile tasks. Each mobile task corresponds to an \gls{osr}, which is composed of a \gls{td} field and a \gls{tr} field. The \gls{td} is used to define the \gls{dl} service requested, the \gls{dl} model to be used and the \gls{dl} target classes, while the \gls{tr} specified the latency and accuracy requirements, the number of \glspl{ue} requested, and the number of jobs (e.g, inference on an image) per second generated by the \glspl{ue}. As shown in Fig.~\ref{fig:framework_oran}-Right, a \gls{td} could be ("Object Recognition", "YOLOX", "\{Person, Car, Bicycle\}"), with the corresponding \gls{tr} defined as ("0.5 s max latency", "0.85 min accuracy", "100 UEs", "50 jobs/sec"). The \gls{td} is submitted to the \gls{sdla} rApp, which is tasked to compute the latency function $l_\tau(\cdot)$ and accuracy function $a_\tau(\cdot)$, which output the latency and accuracy values associated to a given \gls{td}, a given level of task compression and amount of edge resources (see Section \ref{sec:sysmodel} for a more formal definition). The accuracy function can be computed through representative datasets. An initial value of the corresponding latency function can be obtained through emulation.

The latency and accuracy functions are then shared  with the \gls{sesm} xApp running in the \rt. These are ultimately used to solve the \gls{sfesp}, as detailed in Section \ref{sec:semproblem}. The output of the \gls{sfesp} xApp is ultimately three-fold: (i) select which tasks to admit; (ii) their compression level; and (iii) the computational resources (GPU/RAM) and the number of \glspl{prb} assigned to each admitted task. Real-time information about the available computational resources and the current radio-level statistics are provided to the xApp through the E2 interface. The former is used by the \gls{sfesp} to properly account for the resources that are actually available in the RAN edge, which are shared through an Enriched Interface (EI) to the RAN. The latter are used to select and update the appropriate latency function from the \gls{sdla} according to the radio channel status. The radio slicing and computation slicing are respectively shared with the \gls{cu} and the RAN edge through the E2 interface. The \gls{cu} then takes care of propagating the slicing information to the appropriate \glspl{du}. The compression level per task is fed back to the \gls{vno}, which then communicates this information to the \glspl{ue}. We acknowledge that this is impractical at scale, however, as of now,  the O-RAN specifications do not allow for direct communication between \gls{ric} apps and device applications.

\subsection{A Walk-through of \FW.}
We provide a simplified walk-through of an actual slicing request and enforcement operation in \FW on the right side of Fig.~\ref{fig:framework_oran}. First, \glspl{td} are sent to the SDLA rApp (\textbf{Step 1}). If latency/accuracy functions are not already present, they are computed by using the appropriate datasets/models and stored in the \nrt (\textbf{Step 2}). Otherwise, the functions are sent to the SESM xApp (\textbf{Step 3}), which receives the TRs (\textbf{Step 4}) and the current radio/edge status (\textbf{Step 5}), which are used used to produce the RAN and edge slicing (\textbf{Step 6}).  Finally,  the current radio/edge status may be shared with the SDLA rApp for refinement of the latency functions (\textbf{Step 7}) to be used for future slicing decisions. If slice requests change, e.g., because a new task is created, a new slicing allocation is computed. Note that new and already running tasks are equally considered, thus it may happen that previously running tasks are no longer admitted and must be terminated.

\section{Semantic Flexible Edge Slicing (SF-ESP)}\label{sec:semproblem}

We introduce the system model in Section \ref{sec:sysmodel}. Then, we formalize the SF-ESP and prove its NP-hardness in Section \ref{sec:sesp_formulation}. We propose a greedy algorithm in Section \ref{sec:approximate}. 

\subsection{System Model}\label{sec:sysmodel}

\begin{figure}[!t]
  \centering
  \includegraphics[width=\linewidth]{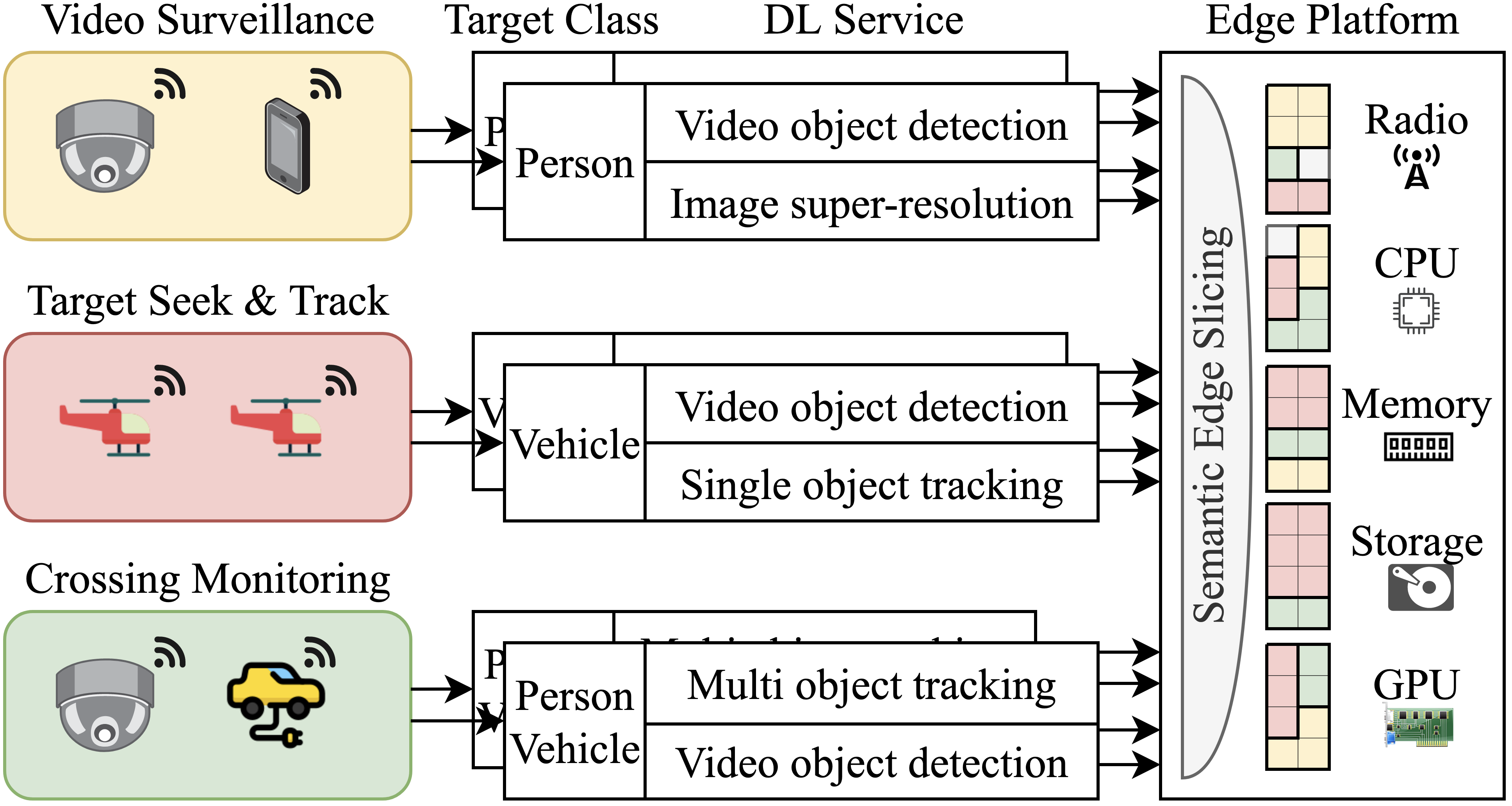}
  \caption{System model example with $C = 3$ application classes, each of which is run by $|D_c|=2, \forall c \in \mathbb{C}$ devices. Each device requests $|T_{cd}|=2, \forall c,d$ tasks to be offloaded to the Edge infrastructure, thus requiring the concurrent allocation of $m=5$ types of radio and compute resources.
  \vspace{-0.5cm}
  }
  \label{fig:sysmodel}
\end{figure}

We define an \textit{application class} as a high-level objective that has to be achieved through the execution of one or more \textit{\gls{dl} tasks} with certain requirements. Every application class specifies the \gls{dl} service, the classes of objects over which the \gls{dl} service is supposed to be applied to, and the requirements for maximum delay and minimum expected accuracy that a device running that application must satisfy. For example, a monitoring application class could require the detection and tracking of person and vehicle objects located in proximity of a road intersection with a minimum expected accuracy of 0.50~mAP and maximum end-to-end delay of 800~ms. Fig~\ref{fig:sysmodel} shows an example with 3 application classes. 

Let $\mathbb{C} = \{1,\ldots,C\}$ be the set containing the application classes. The set of devices running an application class $c \in \mathbb{C}$ is $D_c$. A \textit{device} $d \in D_c$, according to its application class $c$, submits a set of tasks $T_{cd}$ to be offloaded on the RAN edge using its wireless link. A task, uniquely identified at the system level by the tuple $(c,d,t)$, is the periodic execution at the edge of a \gls{dl} service over certain classes of objects, which is applied over a stream of inference data sent by the device, and whose results are then sent back to the requesting device, for a period of time not known a priori. 
To make the notation clearer, let us define $\tau=(c,d,t) \in \mathbb{T}$ as a generic task. Given  $\tau$, we define the compression scaling factor as $z_{\tau}\in (0,1]=\{x\in \mathbb{R} | 0 < x \leq 1\} $ such that the bitrate of the inference data stream is scaled by that factor, i.e. $b_{\tau}^z = z_{\tau} b_{\tau}$, where $b_{\tau}^z$ is the compressed stream and $b_{\tau}$ is the original stream without any applied compression. A higher scaling factor implies higher inference accuracy. A lower scaling factor sacrifices the data quality to decrease the file size, thus requiring lower network bandwidth and improving latency. In our model, we assume that the inference data original stream size is constant and depends on the application class. Furthermore, we assume the compression latency as constant for different scaling factors.

Given the type of edge resource $k \in \mathbb{K} = \{1,\ldots,m\}$, we denote with   $s_{\tau k}$ the amount of resource of type $k$ assigned to each task $\tau \in \mathbb{T}$. Resource types can be networking, e.g., Physical Resource Blocks (PRBs), as well as computational, e.g., GPU time and memory needed to run the \gls{dl} models in the RAN edge. Since edge resources are limited and costly, the total amount of assigned resources of type $k$ cannot exceed the  capacity $S_k, \forall k$. Thus, careful resource allocation is needed to avoid over-provisioning. Since not every resource has the same cost, we define the coefficient $p_k$ as the cost associated with each edge resource type $k$.

The performance requirements are imposed by the related application class. Such  requirements are defined in terms of (i) minimum expected prediction accuracy $A_c$ on the selected object classes, and (ii) maximum expected end-to-end latency $L_c$ for each of the applications running on the mobile devices belonging to class $c$. By defining $a_{\tau}$ and $l_{\tau}$  respectively as the expected accuracy and latency of task $\tau$, an allocation solution is acceptable only if $a_{\tau} < A_c$ and $l_{\tau} > L_c$, $\forall \tau=(c,d,t) \in \mathbb{T}$. Notice that the accuracy and latency are not trivial functions of the slice allocation and compression factor. Specifically, the accuracy depends on the highly nonlinear output of a \gls{dnn}, while the latency has a strong dependency on the radio technology and channel conditions between the \gls{ru} and the \gls{ue}, even when the slice allocation and the compression factor are given. For this reason, integrating a complex mathematical model to account for all of the great numbers of factors involved  (e.g., \gls{snr}, \gls{mcs}, carrier(s) frequency to name a few) would be impractical. Instead, we consider a data-driven approach where the accuracy and latency functions can be constructed through a regression model, keeping the explicit dependencies of the accuracy $a_{\tau}(z): (0,1] \to \mathbb{R^+}$ and latency $l_{\tau}(z,s): (0,1] \times \mathbb{R^+}^m \to \mathbb{R^+}$ functions on the compression scaling factor and resource allocation, and assume that those are given as part of the problem input. In the performance evaluation, we consider latency and accuracy as piecewise functions defined only for the discrete solution values allowed in our experiments.

\begin{table}[h]
\centering
\caption{Table of Symbols}
\begin{tabular}{|c|l|}
\hline
Symbol & Description\tabularnewline
\hline \hline
$\mathbb{C}$ & Set of all application classes \tabularnewline
\hline
$c$ & Application class index \tabularnewline
\hline
$d$ & Mobile device index running an application \tabularnewline
\hline
$t$ & Task index requested by a device \tabularnewline
\hline
$(c,d,t)$ & $t$-th task requested by device $d$ belonging to class $c$ \tabularnewline
\hline
$\tau$ & the generic task identified by the triplet (c,d,t) \tabularnewline
\hline
$\mathbb{T}$ & Set of all tasks $\tau$ of all devices from all classes \tabularnewline
\hline
$\mathbb{K}$ & Set of all Edge resource types \tabularnewline
\hline
$k$ & Edge resource type index \tabularnewline
\hline
$m$ & Total number of resource types \tabularnewline
\hline
$p_{k}$ & Price of the resource type $k$ \tabularnewline
\hline
$x_{\tau}$ & Admission of task $\tau$ \tabularnewline
\hline
$s_{\tau k}$ & Slice allocation of the resource type $k$ for $\tau$ \tabularnewline
\hline
$s_{\tau}$ & Slice allocation vector $(s_{\tau 1},...,s_{\tau m})$ for $\tau$ \tabularnewline
\hline
$a_{\tau}$ & Expected inference accuracy for the task $\tau$ \tabularnewline
\hline
$l_{\tau}$ & Expected E2E latency for the task $\tau$ \tabularnewline
\hline
$A_{c}$ & Minimum accuracy tolerable for class $c$ tasks \tabularnewline
\hline
$L_{c}$ & Maximum latency tolerable for class $c$ tasks \tabularnewline
\hline
$z_{\tau}$ & Compression scaling factor for the task $\tau$\tabularnewline
\hline
$S_{k}$ & Total capacity of type $k$ resource \tabularnewline
\hline
\end{tabular}
\label{tab:notations}
\vspace{-0.4cm}
\end{table}

\subsection{\gls{sfesp} Problem Formulation}\label{sec:sesp_formulation}

We consider the decision variables to be as follows:

\begin{itemize}
    \item 
$\textbf{x} = [x_{\tau}]$, defined as the task admission vector where the generic element, $x_{\tau}$, is a binary  variable indicating whether task $\tau$ is offloaded to the edge or not;
\item
$\textbf{s}=[s_{\tau}]=[(s_{\tau 1},...,s_{\tau m})]$, i.e., the resource allocation matrix;
\item $\textbf{z} = [z_{\tau}]$ defined as the compression scaling factor vector.
\end{itemize}
Note that the data quality is maximum when $z_{\tau} = 1$ and decreases for lower values of $z_{\tau}$. Consequently, the expected inference accuracy $a_{\tau}(z)$ is directly derived from $z_{\tau}$, as it has no dependency from the resource allocation, while the expected latency $l_{\tau}(z,s)$ is a result of the choice of both $z_{\tau}$ and $\{s_{\tau k}\}_{\forall k}$.
The problem formalization according to the system constraints and definitions is given by:


\begin{usecase}[Semantic Flexible Edge Slicing Problem (SF-ESP)\vspace{0.1cm}]
\begin{maxi!}|s|[0]{\textbf{x}, \textbf{s}, \textbf{z}}
{\sum_{\tau \in \mathbb{T}} \sum_{k}^{m} p_k (S_k - s_{\tau k}) x_{\tau} \label{obj}}{\label{opt1}}{}
\addConstraint{\sum_{\tau \in \mathbb{T}} s_{\tau k} x_{\tau}}{\leq S_k,\quad\label{const1.1}}{k=1,\ldots,m}
\addConstraint{z_{\tau}}{\in (0,1],\label{const1.2}}{\forall \tau \in \mathbb{T}}
\addConstraint{a_{\tau}(z_\tau)}{\geq A_{c}x_\tau,\label{const1.3}}{\forall \tau \in \mathbb{T}, c \in \mathbb{C}}
\addConstraint{l_{\tau}(z_\tau,s_\tau)x_\tau}{\leq L_{c},\label{const1.4}}{\forall \tau \in \mathbb{T}, c \in \mathbb{C}}
\addConstraint{x_{\tau}}{\in \{0,1\},\label{const1.5}}{\forall \tau \in \mathbb{T}}.
\end{maxi!}
\end{usecase}

The objective function (\ref{obj}) maximizes the number of allocated tasks $x_\tau$ according to their priority $p_k$, while minimizing the allocated resources $s_{\tau k}$.
Notice that the \gls{sfesp} includes both integer and continuous variables, thus it belongs to the class of mixed integer nonlinear problems (MINLP). Theorem \ref{nphard} below proves that the problem is NP-hard.
\begin{theorem}
  \label{nphard}
  The \gls{sfesp} is NP-hard.
\end{theorem}\begin{proof}
  We  prove the result by showing that  the binary multidimensional Knapsack problem (0/1 d-KP), which is NP-hard \cite{Kellerer2004}, can be reduced to an instance of the \gls{sfesp} in polynomial time.
  Let us assume that the compression factor is fixed to $z_{\tau} = 1, \forall \tau$, and the slice allocation $s_{\tau k}$ is given for every task and resource type. Then let us ignore the constraints on performance by making them always satisfied, i.e., by setting $A_1 = 0$ and $L_1 = \inf$.  The problem now has only \textbf{x} as the decision variable and the value and weight of each task are known and constant. The problem thus is an instance of the 0/1 d-KP, whose statement is the following: given a set of items (tasks), each with a multidimensional weight (resource allocation) and a value (unused resources by their price), determine which items to include in a collection so that the total weight is less than or equal to a given limit (total resources) and the total value is maximized.
  We observe that the \gls{sfesp} is a reduction of 0/1 d-KP that can be built in polynomial time. \vspace{-0.2cm}
\end{proof}
\textbf{The above proof also suggests that \gls{sfesp} is a harder problem than 0/1 d-KP, as it is a combination of the 0/1 d-KP, and a variant of the strongly correlated knapsack with variable weights and  non-linear constraints}. Even though an algorithm with $(1-\epsilon)$-approximation ratio exists for the 0/1 d-KP \cite{FRIEZE1984100}, for the strongly correlated knapsack with variable weights an algorithm with an acceptable approximation ratio is available only for the simpler case where constraints are linear \cite[KLC2]{Nip2007klc}. 
Thus, we provide a greedy heuristic algorithm for which, however, the existing results do not permit to obtain a non-trivial approximation ratio. \vspace{-0.2cm}

\subsection{Greedy Algorithm for the \gls{sfesp}}\label{sec:approximate}

Given the NP-hardness of the \gls{sfesp}, we propose a greedy heuristic to find a sub-optimal solution with low computational complexity, which is based on the primal effective gradient method of \cite{toyoda1975} for the 0/1 d-KP.
This method sorts tasks based on their effective gradients, a measure of the task's relative value according to a penalty vector that prioritizes the allocation of unused resources, then it admits tasks with the highest gradients.
However, to calculate the gradient of a task, we need to first find its resource requirement.
If we assume that the latency function $l_\tau(z_\tau,s_\tau)$ is monotonically increasing over the compression factor $z_\tau$, then the optimal task compression factor $z_\tau^*$ is the minimum that satisfies the accuracy requirement $A_c$ from (\ref{const1.3}):
\begin{equation}
z_\tau^*=\min_{z_\tau} z_\tau\; s.t.\: a_\tau(z_\tau) > A_c
\label{opt:z}
\end{equation}
As for the resource allocation $s_\tau$, our requirement-driven task definition allows for the latency and accuracy constraints to be satisfied with several combinations of resource allocations, with the best being not the minimum, but the one that maximizes the number of admitted tasks. Then, the optimal choice is to balance resource consumption according to resource availability. E.g., if radio resources are scarce, instead of depleting them, we allocate a few radio resources and balance the increased network latency by lowering the processing delay through increased compute resources, to allow additional tasks to still use the remaining radio resources. We achieve this behavior by maximizing the primal gradient function of \cite{toyoda1975}:
\begin{equation}
\begin{aligned}
& s_{\tau}^* = \argmax_{s_{\tau}} PG(s_{\tau}) \\
& s.t.\; l_\tau(z_\tau^*, s_{\tau}) < L_c,\, s_{\tau k} \leq S_k - \left(\sum_{\tau \in \mathbb{T}} s_{\tau k}x_\tau\right), \forall k
\end{aligned}
\label{opt:s}
\end{equation}

To efficiently find a solution to Eqs. \ref{opt:z} and \ref{opt:s}, which depend on the definition of the accuracy and latency functions, it would be necessary to know the properties of the functions (e.g., monotonicity, convexity). Since we consider accuracy and latency as generic functions, we solve the equations through the enumeration of the resource allocation solution space.


\begin{algorithm}[h]
	\caption{Greedy Algorithm for the \gls{sfesp}} 
	\begin{algorithmic}[1]
	    \State $T_c \gets \mathbb{T}$ \label{alg:init_candidate}
	    \Comment consider all tasks candidate for admission
	    \ForAll{$\tau \in \mathbb{T}$}
	        \State $G_\tau \gets 0, x_\tau \gets 0, s_\tau \gets (0,...,0) , z_\tau \gets 1$\label{alg:init_solution}
	        \If{$\exists\, z_\tau^*$}
	            \Comment if minimum accuracy can be met
                \State $z_\tau \gets z_\tau^*$ \label{alg:find_compression}
                \Comment save the optimal compression factor
            \Else
                \State $\mathbb{T}_c \gets \mathbb{T}_c \setminus \tau$  \label{alg:prune_z}
            \EndIf
        \EndFor
	    \Repeat  \label{alg:loop_start}
	        \For{$k \gets 1,m$}
    	        \State $o_k \gets \sum_{\tau \in T} s_{\tau k}x_\tau$ \label{alg:occ_res}
    	        \Comment occupied resources 
	        \EndFor
	        \ForAll{$\tau \in \mathbb{T}_c$}
	            \If{$\exists\, G_{k} \gets \max_{s_{\tau k}} PG(s_{\tau})\; s.t.\: s_{\tau k} \leq S_k - o_k, \forall k$} \label{alg:run_pg}
	                \State $s_{\tau} \gets \argmax_{s_{\tau}} PG(s_{\tau})\; s.t.\: s_{\tau k} \leq S_k - o_k$ \label{alg:s_pg}
	            \Else
	                \State $\mathbb{T}_c \gets \mathbb{T}_c \setminus \tau$ \label{alg:rem_unfeasible}
                \EndIf
	        \EndFor
	        \State $\tau \gets \tau \mid G_\tau = \max\{G_\tau\}_{ \forall \tau}$ \label{alg:max_g}
	        \State $x_\tau \gets 1$ \label{alg:admit}
	        \Comment admit task whose gradient is maximum
            \State $\mathbb{T}_c \gets \mathbb{T}_c \setminus \tau$ \label{alg:rem_admitted}
	    \Until{$T_c = \emptyset$} \label{alg:loop_end}
	    \Ret $(x_\tau,s_\tau,z_\tau)_{ \forall \tau \in \mathbb{T}}$ \label{alg:ret}
	    \Function{PG}{$s_{\tau}$} \label{pg:fn}
	        \Comment calculate the primal gradient
	        \If{$o_k = 0, \forall k$}
	            \Comment penalize resource usage equally
	            \Ret $ (\sum_k^m p_k(S_k-s_{\tau k})) n^{1/2} /(\sum_k^m s_{\tau k}/S_k)$ \label{pg:free_res}
	        \Else
	            \Comment penalize resource usage as per availability
	            \Ret $ (\sum_k^m p_k(S_k-s_{\tau k})) (\sum_k^m o_k^2)^{1/2} /(\sum_k^m s_{\tau k}o_k/S_k)$ \label{pg:ret}
	        \EndIf
	    \EndFunction \label{pg:end_fn}
	\end{algorithmic} 
	\label{greedy}
\end{algorithm}

The preliminary step of the greedy algorithm (Alg.~\ref{greedy}) is to (i) include all submitted tasks to the candidate task set (line \ref{alg:init_candidate}), that contains the tasks that are considered feasible and worth of admission, and (ii) initialize the solution by setting the task admission vector and resource allocation matrix to zero, and the compression scaling factor to the unitary vector (line \ref{alg:init_solution}).
Then, for each task, the optimal compression factor $z^*$ is calculated according to its target accuracy (line \ref{alg:find_compression}), as per Eq.~\ref{opt:z}. An initial pruning of the candidate task set is performed by removing tasks whose target accuracy can not be met for any compression factor (line \ref{alg:prune_z}).
The main loop of the algorithm (lines \ref{alg:loop_start}-\ref{alg:loop_end}) examines the tasks in the candidate task set to find the most convenient one to admit, based on the current resource occupation and until the set empties.
First, the current resource occupation vector is updated (line \ref{alg:occ_res}). After that, the maximum primal gradient of each task in the candidate task set is calculated by exploring the feasible resource allocations (line \ref{alg:run_pg}), following Eq.~ \ref{opt:s}.
The primal gradient is calculated according to the function, defined in lines \ref{pg:fn}-\ref{pg:end_fn}, in which the return value is computed differently whether resources are currently free (line \ref{pg:free_res}) or not (\ref{pg:ret}).
If the maximum gradient is found, then the corresponding resource allocation for the examined task is saved (line \ref{alg:s_pg}), otherwise the task is discarded (line \ref{alg:rem_unfeasible}).
Then, the task with the maximum value of maximum primal gradient is found (line \ref{alg:max_g}), admitted by setting to one its corresponding value of the task admission vector (line \ref{alg:admit}) and therefore removed from the candidate task set (line \ref{alg:rem_admitted}).
Finally, after the loop ends, the task admission vector, the resource allocation matrix, and the scaling factor vector are returned
as the solution of the \gls{sfesp} 
(line \ref{alg:ret}).

\section{Performance Evaluation}\label{sec:experiments}

We evaluate the performance of \FW through extensive numerical analysis (Section \ref{sec:numres}) and practical experiments on the Colosseum network emulator (Section \ref{sec:colres}). 

\vspace{-0.2cm}

\subsection{Experimental Setup} 


\textbf{Applications and datasets.}~As far as the DL services are concerned, we consider object detection  and instance segmentation, which are state-of-the-art problems in computer vision (CV). For the former, we consider (i)  the widely-known Common Objects in Context (COCO) as the dataset, which is a large-scale image database containing more than 200K labeled examples across 80 object classes \cite{lin2014microsoft}; (ii) the YOLOX  classifier, which is based on the Modified CSP v5 as the backbone and has 54.2M parameters \cite{ge2021yolox}. For the latter, we selected (i) the Cityscapes dataset, which contains pixel-level annotated video sequences of street scenes recorded in 50 different cities \cite{cordts2016cityscapes}; (ii) the BiSeNet v2 real-time classifier, which is based on a bilateral segmentation backbone network and has 14.8M parameters \cite{Yu2021}. For performance evaluation purposes, we define a set of 10 object detection tasks in Tab.~\ref{tab:applications}.

\begin{table}[!h]
    \centering
        \caption{Multi-object detection applications.}
    \begin{tabular}{|p{0.26\linewidth} | p{0.58\linewidth}|}
    \hline
        \textbf{Application} & \textbf{Target Classes} \\\hline\hline
        COCO All & Entire set of classes (80) of COCO \\\hline 
        COCO Urban & Bicycle, car, motorcycle, bus, truck, traffic light, stop sign, person\\\hline
        COCO Bags & Handbag, backpack, suitcase \\\hline
        COCO Animals & Bird, cat, dog, horse, sheep, cow, elephant, bear, zebra, giraffe\\\hline
        COCO Person & Person \\\hline
        Cityscapes All & All evaluation classes (19) of Cityscapes\\\hline
        Cityscapes Vehicles & Car, truck, bus, train, motorcycle, bicycle \\\hline
        Cityscapes Objects & Pole, traffic light, traffic sign \\\hline
        Cityscapes Flat & Road, sidewalk \\\hline
        Cityscapes Person & Person \\\hline
    \end{tabular}
    \label{tab:applications}
\vspace{-0.2cm}
\end{table}

\textbf{Baselines.}~For comparison purposes, we consider the following baselines: (1) Sl-EDGE \cite{d2020sl}, the state-of-the-art algorithm for RAN edge slicing; (2) MinRes-SEM, an algorithm that considers the semantics but, instead of flexibly allocating resources, it allocates the minimum resources for each task; (3) FlexRes-N-SEM, which implements flexible resource allocation following Eq.~ \ref{opt:s} but does not consider the semantics; (4) HighComp, which compresses each task to 10\% of its original size, so as to reach mAP of about 0.25 in the COCO dataset. This is a baseline that tries to compress aggressively tasks to minimize resources; (5) HighRes, which statically allocates tasks 20\% of the total amount of resources. This is a baseline that attempts to maximize the probability that admitted tasks will meet application constraints.\vspace{0.1cm}

\textbf{Prototype on Colosseum.}~We designed and developed a proof of concept of \FW on the Colosseum network emulator \cite{bonati2021colosseum}, and used the open-source SCOPE framework \cite{bonati2021scope} as prototyping platform for NextG
systems. Since SCOPE did not support the uplink slicing of resources, we extended SCOPE to implement uplink slicing as well. 
\vspace{-0.2cm}


\begin{figure}[!h]
    \centering
    \includegraphics[width=\columnwidth]{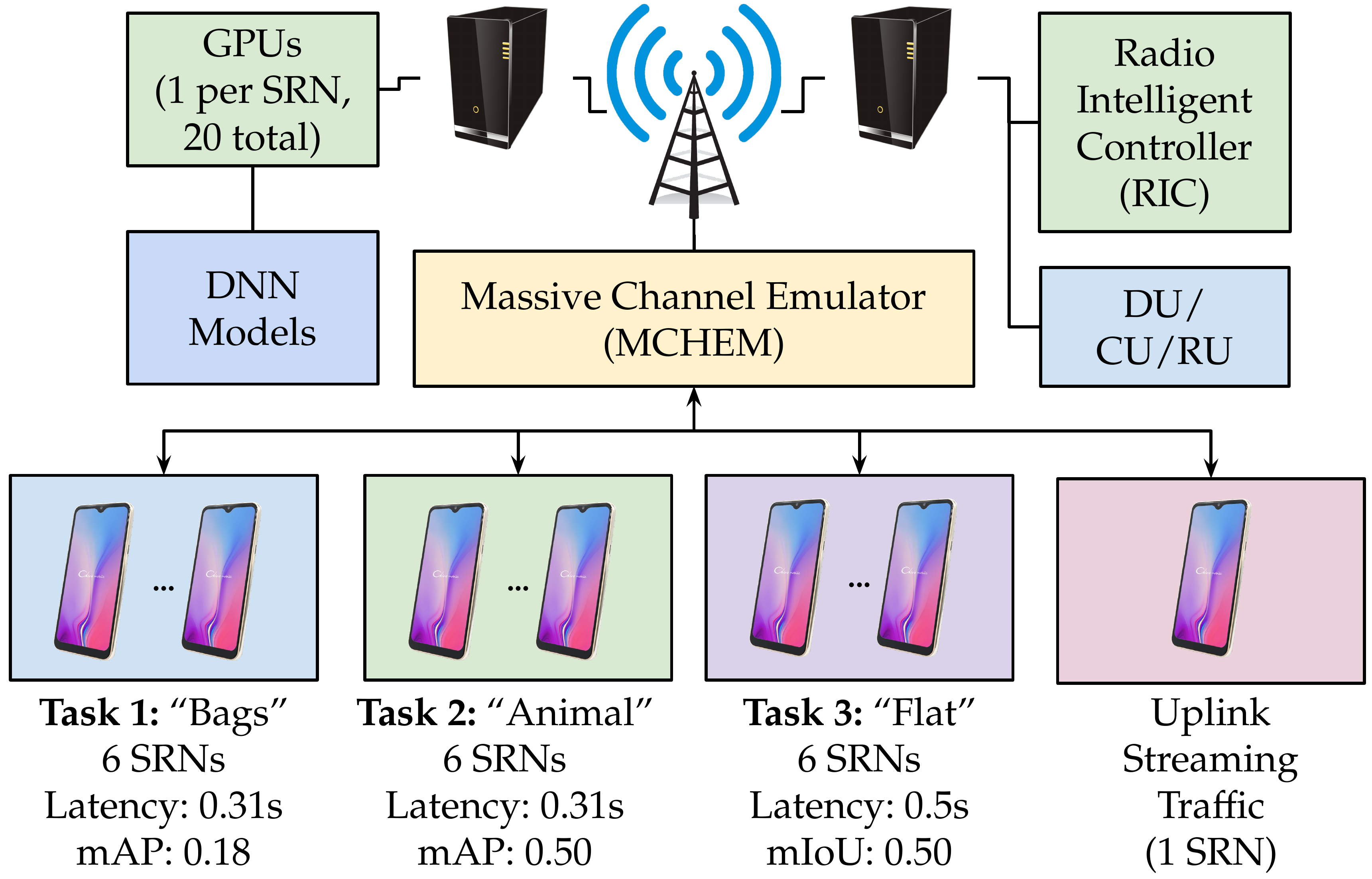}
    \caption{Experimental setup on Colosseum.\vspace{-0.3cm}}
    \label{fig:colosseum_setup}
\end{figure}


Fig.~\ref{fig:colosseum_setup} shows a high-level overview of the \FW prototype.  We utilize a set of 20 \glspl{srn} to implement the O-RAN network, with 1 \gls{srn} used to process received jobs of admitted tasks and to implement the DU/CU/RU and the \gls{ric}, where we run the slice admission system and the solvers of the \gls{sfesp}, implemented in MATLAB. Out of the remaining 19 \glspl{srn}, to emulate traffic separated from the mobile applications requiring RAN slices, we use one \gls{srn} to generate uplink streaming traffic with the \textit{iperf} tool. The other 18 \glspl{srn} are used to implement a system where a \gls{vno} requests three slices for object detection tasks. Up to 20 Tesla K40m GPUs can be utilized to run the \glspl{dnn}. As for the PHY, we utilize the standard SCOPE parameters, i.e., 10 MHz of bandwidth corresponding to 50 \glspl{prb} in total grouped in 17 \glspl{rbg}. We assign the uplink streaming traffic 2 \glspl{rbg}, thus, 15 \glspl{rbg} are available for slicing.

\subsection{Numerical Results}\label{sec:numres}

\begin{figure*}[!h]
  \centering
  \begin{subfigure}[]{\textwidth}
    \centering
    \includegraphics[height=22pt]{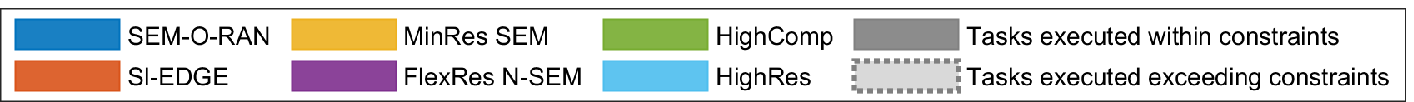}
  \end{subfigure}
    \begin{subfigure}[]{0.49\textwidth}
      \centering
      \includegraphics[width=\textwidth]{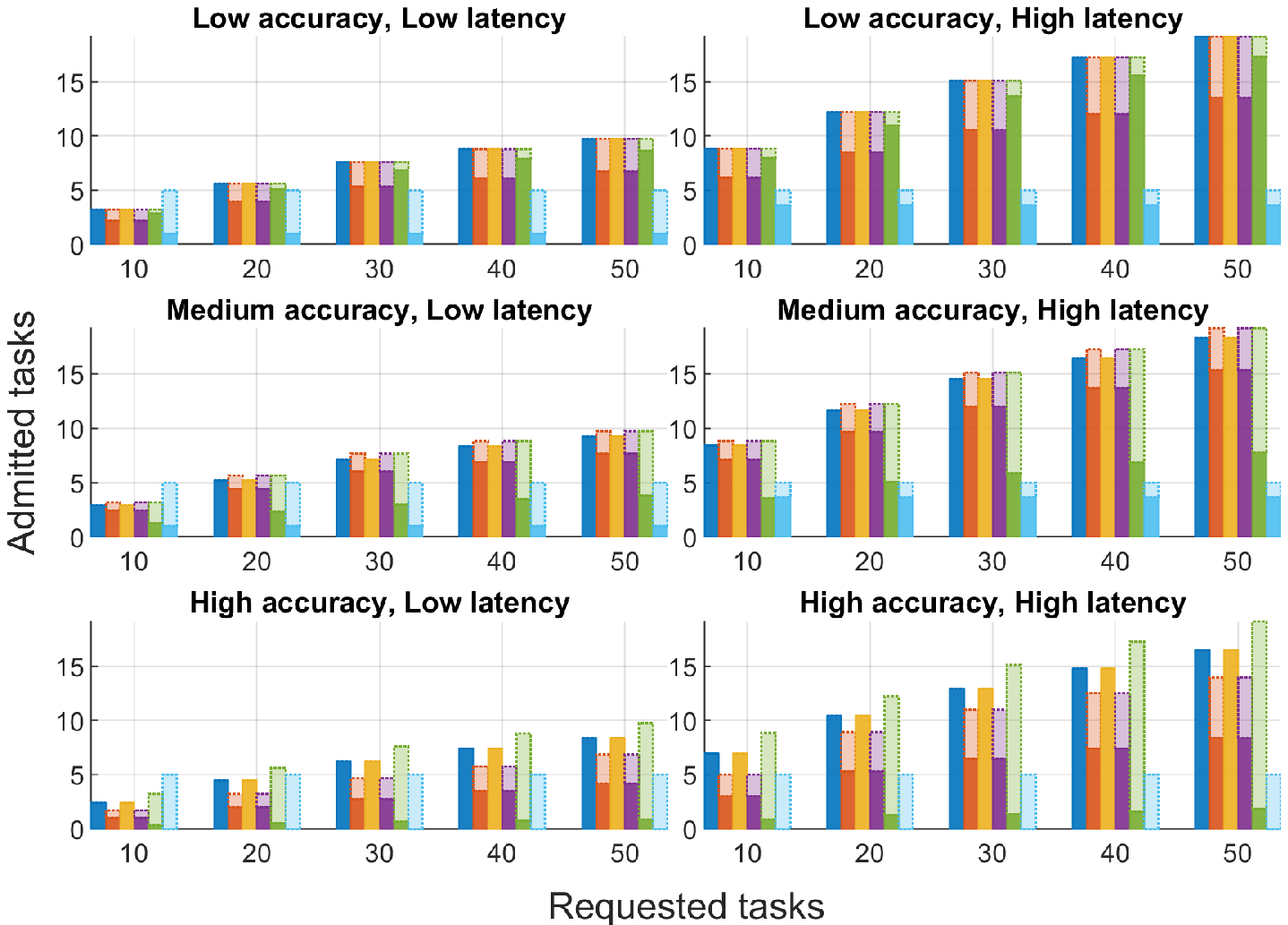}
      \caption{Numerical results with 2 types of edge/network resources.}
      \label{}
  \end{subfigure}
    \hfill
  \begin{subfigure}[]{0.49\textwidth}
      \centering
      \includegraphics[width=\textwidth]{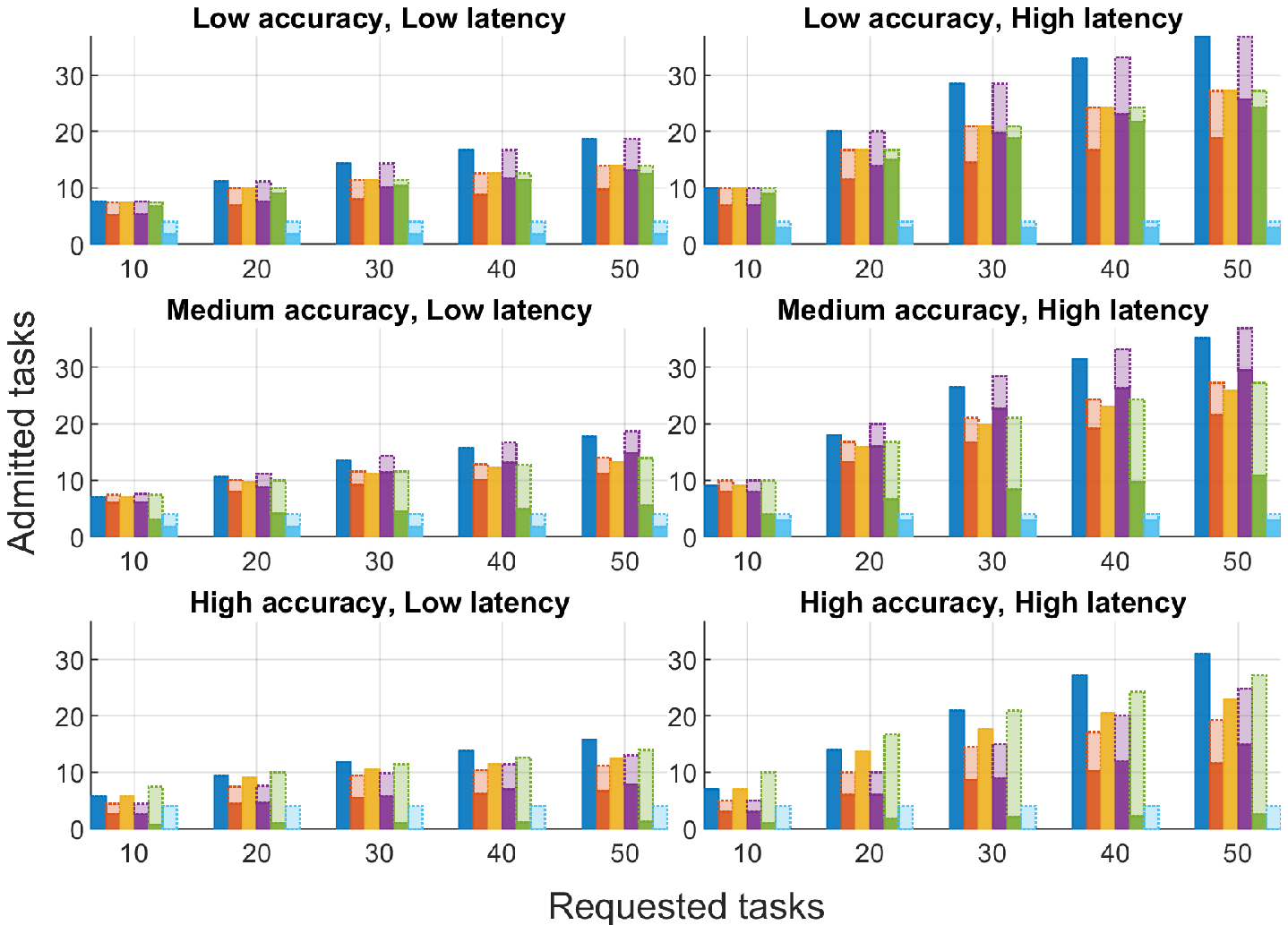}
      \caption{Numerical results with 4 types of edge/network  resources.}
      \label{}
  \end{subfigure}
\caption{Numerical results and comparison between \FW and baselines.\vspace{-0.3cm} }
\label{fig:numeric}
\end{figure*}

Fig.~\ref{fig:numeric} shows the number of allocated tasks by \FW and the baseline algorithms, as a function of the number of requested tasks. To further investigate the impact of our approach, we consider (i) different numbers (2 and 4) of edge/network resources (e.g., CPUs, GPUs, PRBs, etc.); (ii) different thresholds of accuracy (``low'', ``medium'' and ``high'') and latency (``low'', ``high''). We define the accuracy thresholds $A_c$ as 0.20, 0.35, and 0.55 mAP for object detection tasks and 0.35, 0.50, and 0.70 mean Intersection over Union (mIoU) for instance segmentation tasks, while for latency threshold $L_c$ we choose 0.2 seconds and 0.7 seconds. Tasks are equally distributed across the applications defined in Tab.~\ref{tab:applications}. We empirically formulate a latency function $l_\tau$ that expresses the computational and network latency as a function of compression factor, resource allocation, and task generation rate. 

Fig.~\ref{fig:numeric}(a) shows that, in general, the performance of \FW is similar to the one given by MinRes-SEM. Even when the requirements are medium accuracy and high latency, \FW allocates 20\% more tasks than Sl-EDGE and FleRes-N-SEM, and 402\% more tasks than HighRes, when 50 tasks are generated. On the other hand, when the accuracy requirements deviate from medium, we start to notice that \FW delivers significantly better performance than Sl-EDGE. Specifically, we notice that when high mAP/mIoU is required, only \FW and MinRes-SEM are able to allocate tasks that meet the requirements. Sl-EDGE does not allocate tasks since Sl-EDGE considers all the tasks as belonging to the "All" application, which can never reach the required mAP/mIoU of 0.55/0.70 (see the left side of Fig.~\ref{fig:semantic_allocation}). While HighComp and HighRes do allocate tasks, they will not meet the requirements. The reason is that HighComp and HighRes allocate tasks while being agnostic of the target latency and accuracy. The effect of joint semantic slicing and flexible resource allocation is even more evident in Fig.~\ref{fig:numeric}(b), where more types of edge/network resources are considered. In this case, \FW overperforms all the other schemes in all of the considered scenarios, especially when the number of tasks increases and the requirements become more stringent. \textbf{The results indicate that \FW allocates up to 169\% more tasks than the existing state-of-the-art Sl-EDGE algorithm and 18,5\% on average.}
\vspace{-0.2cm}

\subsection{Experimental  Results}\label{sec:colres}

\begin{figure*}[!h]
  \centering
  \begin{subfigure}[]{\textwidth}
    \centering
    \includegraphics[height=14pt]{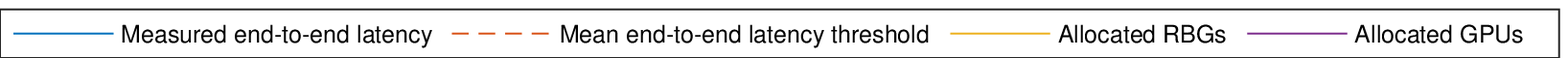}
  \end{subfigure}
  \begin{subfigure}[]{0.325\textwidth}
    \centering
    \includegraphics[width=\textwidth]{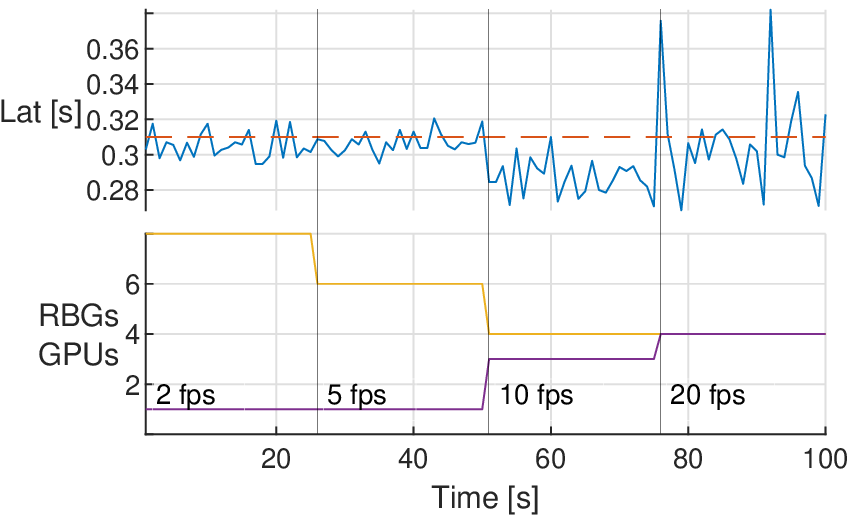}
    \caption{SEM-O-RAN, "Bags", $z=0.28$}
    \label{}
  \end{subfigure}
  \hfill
  \begin{subfigure}[]{0.325\textwidth}
    \centering
    \includegraphics[width=\textwidth]{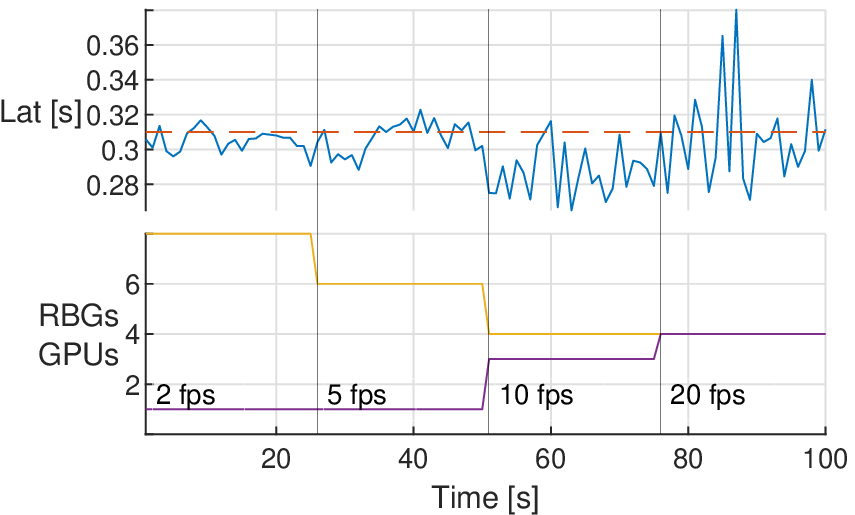}
    \caption{MinRes SEM, "Bags", $z=0.28$}
    \label{}
  \end{subfigure}
  \hfill
  \begin{subfigure}[]{0.325\textwidth}
    \centering
    \includegraphics[width=\textwidth]{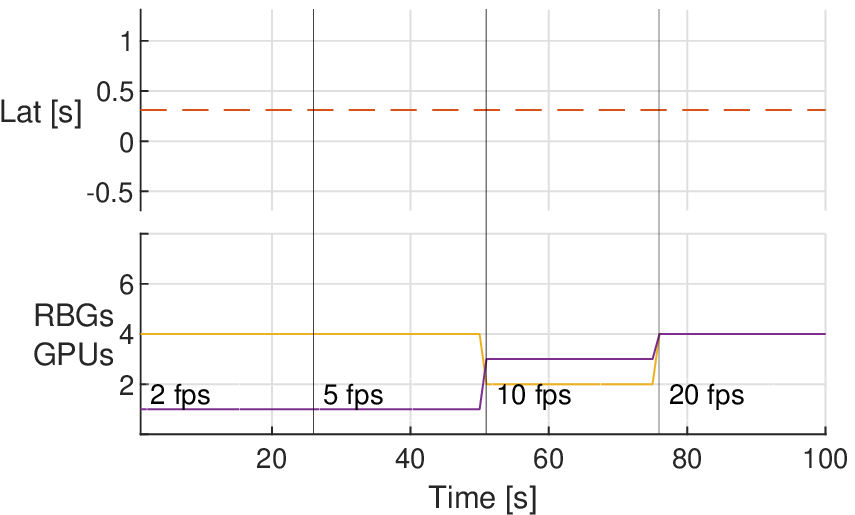}
    \caption{FlexRes-N-SEM, "Bags", {\color{red}$z=0.14$}}
    \label{}
  \end{subfigure}
  \begin{subfigure}[]{0.325\textwidth}
    \centering
    \includegraphics[width=\textwidth]{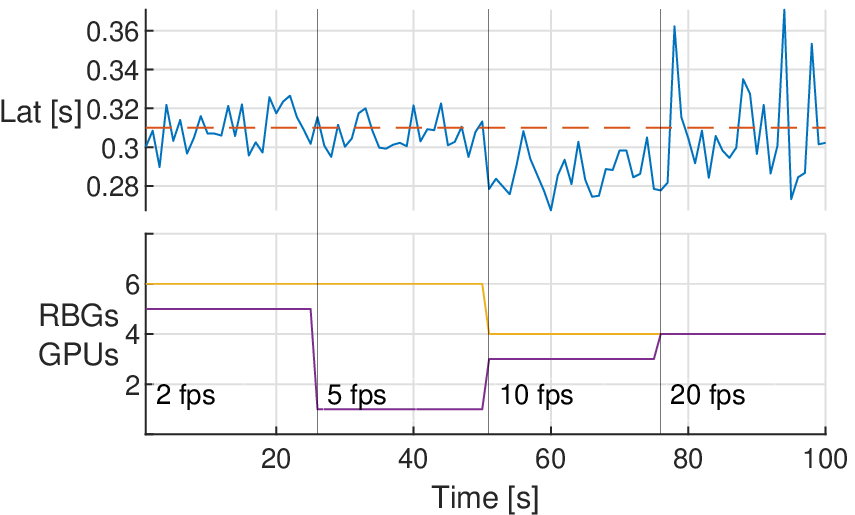}
    \caption{SEM-O-RAN, "Animals", $z=0.28$}
    \label{}
  \end{subfigure}
  \hfill
  \begin{subfigure}[]{0.325\textwidth}
    \centering
    \includegraphics[width=\textwidth]{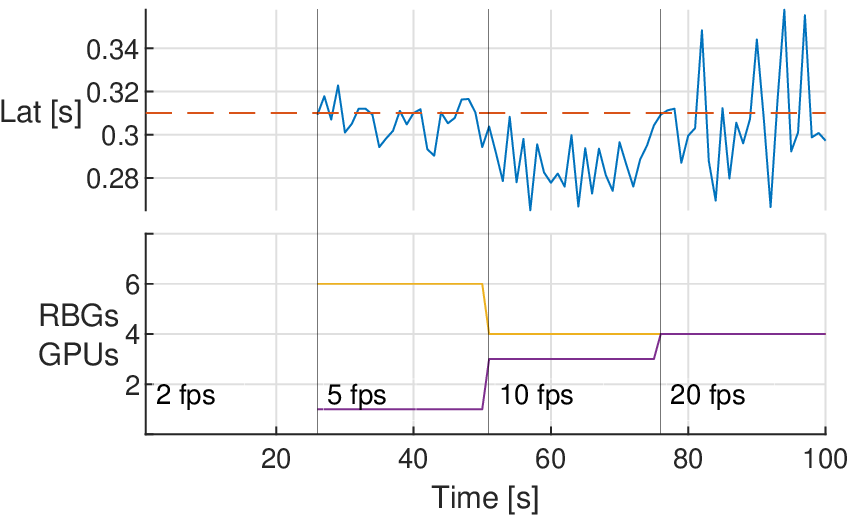}
    \caption{MinRes-SEM, "Animals", $z=0.28$}
    \label{}
  \end{subfigure}
  \hfill
  \begin{subfigure}[]{0.325\textwidth}
    \centering
    \includegraphics[width=\textwidth]{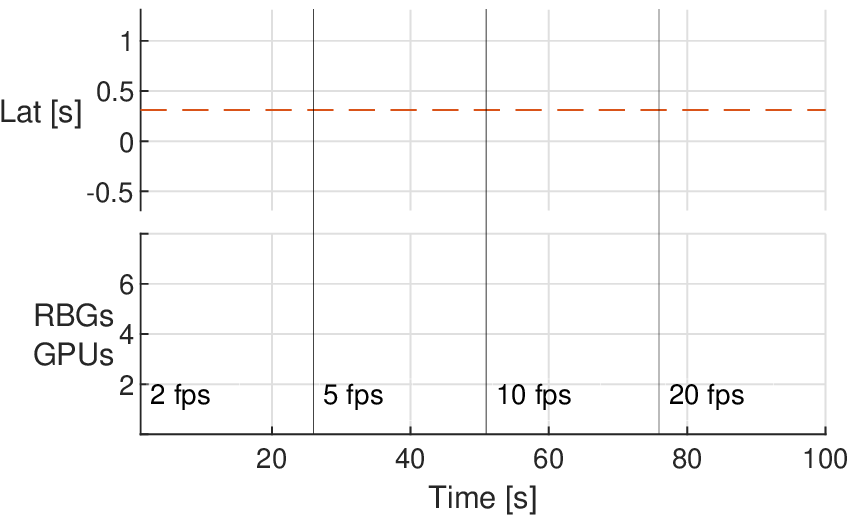}
    \caption{FlexRes-N-SEM, "Animals", {\color{red}$z>1$}}
    \label{}
  \end{subfigure}
  \begin{subfigure}[]{0.325\textwidth}
      \centering
      \includegraphics[width=\textwidth]{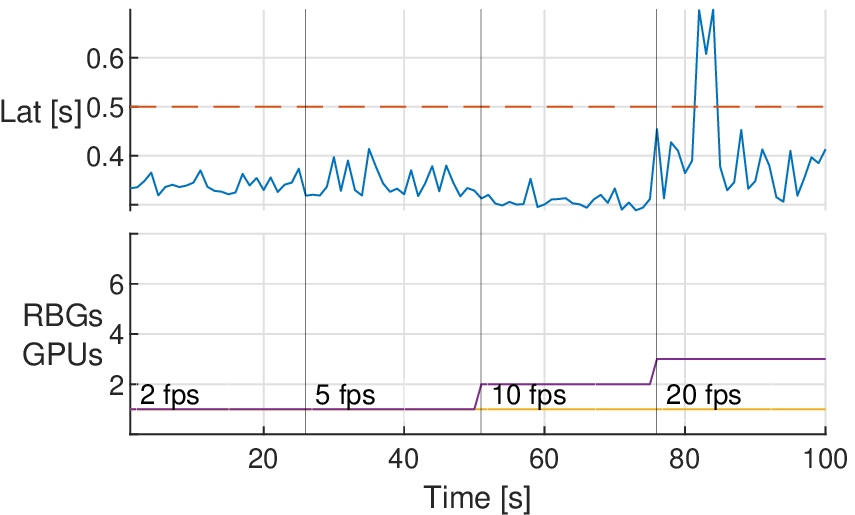}
      \caption{SEM-O-RAN, "Flat", $z=0.08$}
      \label{}
  \end{subfigure}
  \hfill
  \begin{subfigure}[]{0.325\textwidth}
    \centering
    \includegraphics[width=\textwidth]{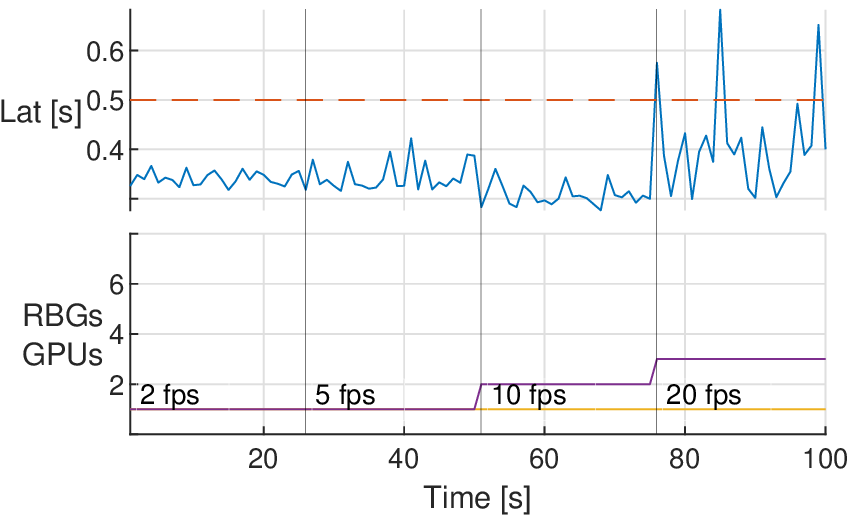}
    \caption{MinRes-SEM, "Flat", $z=0.08$}
    \label{}
  \end{subfigure}
  \hfill
  \begin{subfigure}[]{0.325\textwidth}
    \centering
    \includegraphics[width=\textwidth]{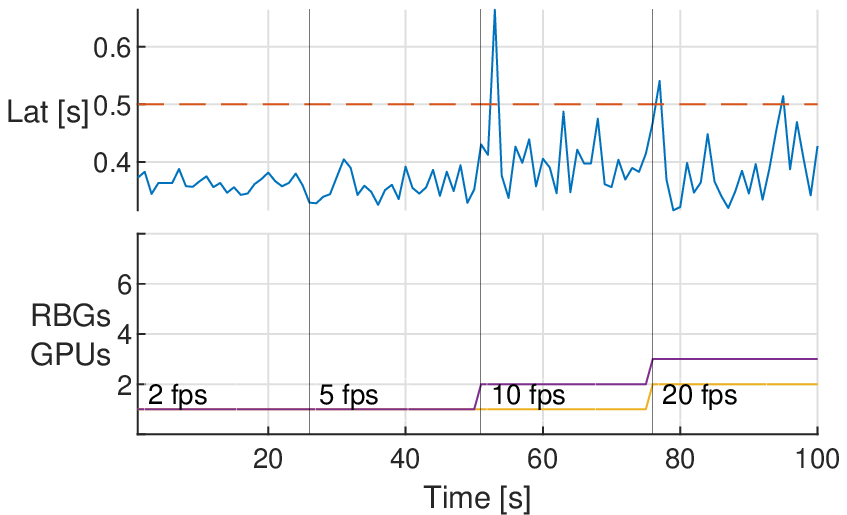}
    \caption{FlexRes-N-SEM, "Flat", $z=0.18$}
    \label{}
  \end{subfigure}
\caption{Experimental results obtained through Colosseum, where we report the end-to-end latency as a function of time, as well as the end-to-end latency threshold requirement. We change the slice requirements by updating the number of generated frames per second (fps) by each UE every period of 25 seconds and show the related output of the slicing algorithm in terms of RBGs (radio resources) and GPUs (computing resources). In each caption, we show the chosen compression rate.\vspace{-0.3cm}}
\label{fig:colosseum}
\end{figure*}

Fig. \ref{fig:colosseum} shows our experimental results on Colosseum, in which we change the VNO slice requirements by updating the number of frames per second (fps) that will be generated by each UE every 25\,seconds, while latency and accuracy constraints are kept constant (values in Fig.~\ref{fig:colosseum_setup}). Whenever the requirements are updated, \gls{sesm} computes a new solution and enforces new slice configurations. Thus, we report the experimental end-to-end latency for each slice as a function of time, as well as the end-to-end latency threshold requirement for each task. To further investigate the advantage of flexible allocation and semantic slicing, we compare \FW to MinRes-SEM  and FlexRes-N-SEM. Accordingly, we show the related output of the slicing algorithm in terms of RBGs (radio resources) and GPUs (computing resources). 

We see that \FW successfully allocates "Bags", "Animals" and "Flat". Notice that the reason why RBG allocation decreases as the fps request decreases is that for lower values of fps, the experienced latency increases, since some time is spent for LTE uplink scheduling requests from the UEs \cite{pocovi2019}. With higher fps, the UE is able to use RBGs granted by the eNB to exchange traffic pertaining to multiple frames, thus leading to lower latency even if  network utilization is higher. In the third and fourth periods, all three tasks are allocated by \FW. The impact of flexible resources is demonstrated in (e) where we see that MinRes-SEM does not allocate "Animals" in the first period. The reason is that \FW is balancing RBGs with GPUs, requesting 6 RBGs and 5 GPUs during the first period. Since MinRes-SEM would have requested 8 RBGs and 1 GPUs, this would have led to 16 RBGs in total, which exceeds system capacity.

Finally, from Figs. (c), (f) and (i) it emerges that FlexRes-N-SEM, by not considering the semantics, performs worse than the former two approaches. By keeping in mind that FlexRes-N-SEM assumes that every task is of type "All", it will compress the tasks in "Bags" to 14\% of their original size to maximize the number of tasks allocated. Conversely, \FW  and MinRes-SEM compress "Bags" to 28\%, which leads to successful allocation since the mAP constraint will be met. Worse yet,  FlexRes-N-SEM will allocate resources for "Bags" but the tasks will fail because they will not meet the required mAP. Thus, even if  FlexRes-N-SEM saves resources by compressing more, it cannot achieve the required mAP. As shown in Fig.~\ref{fig:colosseum}(f), the "Animals" task is never admitted by FlexRes-N-SEM, because it assumes that a mAP of 0.5 can never be reached by "All", while \FW and MinRes-SEM, by considering the semantics, compress the tasks to the optimal level and can successfully admit it. 
As for "Flat", FlexRes-N-SEM is always able to allocate it successfully but, by assuming the type as the more complex "All", it does not select the same aggressive compression factor that instead is chosen by \FW and MinRes-SEM (18\% instead of 8\%), at the cost of higher RBGs consumption in the latest period of Fig.~\ref{fig:colosseum}(i).
\vspace{-0.2cm}

\section{Related work}\label{sec:rw}
\vspace{-0.1cm}
\gls{ran} slicing has attracted significant attention over the last years \cite{d2019slice,mandelli2019satisfying,d2020sl}. Moreover, as the \gls{ran} gets softwarized, \gls{mec} becomes crucial to address the ever-stringent latency demands of mobile applications~\cite{wang2019computation,zhang2019task}. We refer the interested reader to the surveys~\cite{afolabi2018network,wijethilaka2021survey}.


Specific to the slicing of edge resources, Van Huynh \textit{et al.} \cite{van2019optimal} presented a mechanism for slicing of computation, networking, and storage through a deep dueling neural network that provides slices admission while avoiding over-provisioning and maximizing the \gls{vno}'s reward. However, the authors in \cite{van2019optimal} do not  focus on how to partition the \gls{mec} resources and only focus on admission control. Conversely, Ndikumana \textit{et al.}~\cite{ndikumana2019joint} consider the allocation of heterogeneous resources for MEC task offloading, while in \cite{liu2019direct} Liu \emph{et al.} propose a framework for MEC-enabled wireless networks called DIRECT, which however does not consider the case when MEC and networking resources are on the same edge node. Moreover, these frameworks are not \gls{oran}-compatible, which is instead one of the primary targets of this paper.

So far, most of the research focus in O-RAN  has been on designing algorithms for RAN control and optimization. 
Bonati \emph{et al.} \cite{bonati2021intelligence} have developed an xApp running \gls{drl} agents to select the best-performing scheduling policy for each \gls{ran} slice. In our work, we do not select scheduling policies but instead focus on \gls{ran} slicing. D'Oro \emph{et al.} \cite{doro2022orchestran} proposed an orchestration mechanism to select the optimal \gls{dl} models and execution location for each model complying with timescale requirements, resource, and data availability. Conversely, we focus on properly slicing \gls{mec} resources for timely execution of \gls{cv}-based \gls{dl} models under strict accuracy constraints. Although flexible resource allocation has been considered in the context of Virtual Network Function (VNF) \cite{golkarifard2021dynamic,martin2021kpi}, existing formulations do not consider application semantics, and, in general, cannot be easily applied to address edge task offloading problems.

The closest work to ours is Sl-EDGE \cite{d2020sl}, a MEC slicing framework that allows network operators to instantiate heterogeneous edge slices. The key limitation of Sl-EDGE is that it does not consider DL semantics, which is the core advantage of our approach. 
\vspace{-0.2cm}

\section{Concluding Remarks}
\vspace{-0.1cm}
We have proposed \FW, the first \textit{semantics-based} slicing framework for NextG O-RAN networks. \FW delivers better performance  by  semantically compressing the images sent to the edge. Moreover, unlike prior art, \FW does not consider each task as monolithic, but flexibly allocates radio and computational resources so as to maximize the number of admitted tasks.  
We have mathematically formulated the Semantic Flexible Edge Slicing Problem (SF-ESP), demonstrated that it is NP-hard, and proposed a greedy approximation algorithm to solve it efficiently. We have evaluated the performance of \FW through extensive numerical analysis by comparing it to several baseline algorithms including the state-of-the-art scheme \cite{d2020sl}. We have implemented a  prototype of \FW by using the Colosseum network emulator through the SCOPE framework for NextG systems \cite{bonati2021scope}. Our numerical and experimental results show that through our semantic-based approach, \FW improves the number of allocated tasks by up to 169\% with respect to the existing state-of-the-art work, while still meeting accuracy and delay constraints. We believe that beyond the results presented in this paper, the proposed semantic-based approach can serve as the foundation for future research on the utilization of application-level features in the low-level design and optimization of wireless networks. 
\vspace{-0.1cm}
\section*{Acknowledgment of Support and Disclaimer}
\vspace{-0.2cm}
This work is funded in part by the National Science Foundation (NSF) grant CNS-2134973 and CNS-2120447, by an effort sponsored by the U.S. Government under Other Transaction number FA8750-21-9-9000 between SOSSEC, Inc. and the Government., and  by the European Union’s NextGenerationEU instrument, under the Italian National Recovery and Resilience Plan (NRRP), Mission 4 Component 2 Investment 1.3, enlarged partnership “Telecommunications of the Future” (PE0000001), program “RESTART”. 
The U.S. Government is  authorized to reproduce and distribute reprints for Governmental purposes notwithstanding any copyright  notation thereon. The views and conclusions contained herein are those of the authors and should not be interpreted as necessarily representing the official policies or endorsements, either expressed or implied, of the NSF, the Air Force Research Laboratory, the U.S. Government, or SOSSEC, Inc. 
\bibliographystyle{ieeetr}
\bibliography{main}
\end{document}